\documentclass[twocolumn]{aastex62}
\usepackage{appendix}
\usepackage{natbib}
\usepackage{amsmath}
\usepackage{topcapt}
\usepackage{booktabs}
\usepackage{amsmath}
\usepackage{multirow}
\usepackage{longtable}
\usepackage{color,subfigure,lineno}
\usepackage{hyperref}
\newcommand{\RNum}[1]{\uppercase\expandafter{\romannumeral #1\relax}}
\usepackage{rotating}

\newcommand{\hbeta}{H{$\beta$}}

\newcommand{\lya}{Ly\,$\alpha$}
\newcommand{\CIV}{C\,{\sevenrm IV}}

\def\HeII{He\,{\sc ii}}

\def \OIII {[O\,{\sc iii}]}

   \font\sevenrm=cmr7 scaled 1000
\newcommand{\comments}[1]{}

\begin{document}

\title{Dynamical Modeling of the Broad-Line Region with High-Mass Active Galactic Nuclei and Constraints on the Virial Factor}

\author[0000-0002-2052-6400]{Shu Wang}
\affiliation{Astronomy Program, Department of Physics and Astronomy, Seoul National University, Seoul, 08826, Republic of Korea; jhwoo@snu.ac.kr}

\author[0000-0002-8055-5465]{Jong-Hak Woo}
\affiliation{Astronomy Program, Department of Physics and Astronomy, Seoul National University, Seoul, 08826, Republic of Korea; jhwoo@snu.ac.kr}

\author[0000-0002-1961-6361]{Lizvette Villafa\~{n}a}
\affil{Physics Department, California Polytechnic State University, San Luis Obispo, CA 93407, USA}
\affil{Department of Physics and Astronomy, University of California, Los Angeles, CA 90095-1547, USA}

\author[0000-0002-8460-0390]{Tommaso Treu}
\affil{Department of Physics and Astronomy, University of California, Los Angeles, CA 90095-1547, USA}

\author[0000-0001-5802-6041]{Elena Gallo} 
\affil{Department of Astronomy, University of Michigan, Ann Arbor, MI 48109, USA}

\begin{abstract}
We present the results of broad-line region (BLR) dynamical modeling for eight
high-mass active galactic nuclei (AGNs) from the Seoul National University AGN Monitoring Project, by constraining BLR geometry and kinematics as well as black hole (BH) mass ($M_{\rm BH}$). 
We find that the H\(\beta\)-emitting BLRs 
are best described as thick disks viewed at intermediate inclinations, with emission preferentially originating from the far side of the BLR. 
BLR kinematics show a combination of rotational, inflowing and outflowing components. 
By comparing the $M_{\rm BH}$ from dynamical modeling with the virial products based on reverberation lags and line widths, we determine the virial factor $f$ for individual AGNs. Combining our sample with those $M_{\rm BH}$ consistently determined from BLR dynamical modeling, yielding a total of 38 objects, we derive a virial factor for future $M_{\rm BH}$ estimation of log$_{10}({f})_{\rm pred}=0.69\pm0.21$  based on $\sigma_{\rm line,rms}$ and $-0.08\pm0.23$ based on FWHM$_{\rm mean}$.
The derived virial factor is consistent with that inferred by aligning the reverberation-mapped AGNs with quiescent galaxies in the \(M_{\rm BH}\)–\(\sigma_*\) relation,
supporting the assumption that local active and inactive galaxies follow the same \(M_{\rm BH}\)–\(\sigma_*\) relation. 
Our updated $f$ values exhibit an intrinsic dispersion  of  $\sim0.2$ dex, which allows for a more precise $M_{\rm BH}$ estimates than those based on the \(M_{\rm BH}\)–\(\sigma_*\) relation. Our sample extends the dynamical modeling-based reverberation sample to \(M_{\rm BH} \sim [10^8, 10^{8.5}] M_{\odot}\) range, where the virial factor from the the AGN \(M_{\rm BH}\)–\(\sigma_*\) relation remains poorly constrained, underscoring the unique value of dynamical modeling analysis in constraining the $M_{\rm BH}$ of the most massive BHs.

\end{abstract}


\section{Introduction}

The discovery of a strong correlation between the black hole (BH) mass ($M_{\rm BH}$) and the bulge stellar velocity dispersion ($\sigma_*$) is a cornerstone of our understanding of galaxy evolution  \citep[e.g.][]{Ferrarese00,Gebhardt00, Gultekin09,Woo10,Bennert11,Bennert15,McConnell_Ma13}.  It suggests a profound coevolutionary link between supermassive BHs and their host galaxies \citep[e.g.,][]{Kormendy13} established via AGN feedback \citep[e.g.,][]{Kauffmann00,Volonteri03} and/or hierarchical growth driven by mergers \citep[e.g.,][]{Peng07,Jahnke11}.  Tracing the cosmic evolution of BH-galaxy correlations is crucial for understanding the underlined physics \citep[e.g.,][]{Woo06,Woo08,Treu07,Sexton19, Ding21,Zhuang23,Pacucci24}.

Our understanding of the $M_{\rm BH}$--$\sigma_*$ relation is primarily established based on local quiescent galaxies, where $M_{\rm BH}$ can be accurately measured using spatially resolved kinematics of stars, gas, or megamasers. However, extending such measurements to higher redshifts is extremely challenging, as the BH's sphere of influence becomes too small to be resolved with current facilities. Active galactic nuclei (AGNs) provide a valuable alternative, as their $M_{\rm BH}$ can be estimated via reverberation mapping \citep[RM;][]{Blandford_McKee_1982,Peterson93}. 

In RM, the size of broad-line region (BLR; $R_{\rm BLR}$) is inferred from the time delay ($\tau$) between the continuum variations and the responses of the broad emission lines flux. The $R_{\rm BLR}$ is then combined with the velocity width ($\Delta V$) of broad emission lines  to give a virial $M_{\rm BH}$ measurement: $M_{\mathrm{BH}} = f \, \frac{R_{\rm BLR} \;\Delta V^2}{G}$, where the term $\frac{R_{\rm BLR} \;\Delta V^2}{G}$ is called the virial product with the unit of mass, and $f$ is a scaling factor accounting for the BLR orientation, geometry and kinematics. One of the key findings in RM studies is the discovery of a tight relation between BLR radius and AGN luminosity  \citep[e.g.,][]{Kaspi00,Bentz09,Bentz13,Grier17b,Du19,Fonseca-Alvarez20,Shen24,Woo24,Wang24}. This radius-luminosity relation enables BH mass estimation based on single-epoch (SE) spectrum which greatly enhances our understanding of SMBH demographics across different redshifts \citep[e.g.,][]{Greene07,Vestergaard09,Shen11,Kelly13,Rakshit19, Wu-Q22, Wu22,Cho24, Pan25}. 

Due to our limited knowledge of BLR properties, the classical RM that measures the average BLR size cannot determine $f$ for individual AGNs. Instead, a single sample-averaged $f$ is commonly adopted, determined by matching RM-based AGN $M_{\rm BH}$ to the $M_{\mathrm{BH}}-\sigma_*$ relation of quiescent galaxies \citep[e.g.,][]{Onken04,Collin06,Woo10,Woo13,Woo15,Park12a,Grier13b,Hokim14,Hokim15,Batiste17,Yang24}, with the assumption that active and inactivate galaxies follow the same relation. Under this framework,  \citet{Woo15} estimated the logarithmic virial factor as log$f=0.65\pm0.12$ based on line velocity dispersion ($\sigma_{\rm line}$) and  log$f=0.05\pm0.12$ based on full-width-at-half-maximum (FWHM). In addition to this uncertainty, the $\sim 0.4$ dex intrinsic scatter of AGN $M_{\mathrm{BH}}-\sigma_*$ relation \citep[e.g.,][]{Woo15} propagates to the virial factor, becoming the primary sources of the systematic uncertainty in RM-based $M_{\rm BH}$.

To reduce this systematic uncertainty, additional constraints on the BLR structure and dynamics are required. Velocity-resolved RM analysis, which measures time lags across different line-of-sight velocity bins, provides useful qualitative constraints on BLR kinematics \citep[e.g.,][]{Gaskell88,Crenshaw90, Bentz09b,Denney10, Grier13a,Du16, Lu21, Feng21b, Feng24, Feng25, U22, Bao22, Zastrocky24,Fries24, Wang25}. Based on $\sim 94$ AGNs with velocity-resolved lags, \citet{Wang25} shows that while the most common velocity-resolved structure is consistent with disk-like rotation, BLRs do exhibit diverse kinematics including inflow and outflow. A more detailed approach is to reconstruct the two-dimensional transfer function \citep[e.g.,][]{Peterson93,Horne04,Skielboe15}, which describes how the line responsitivity varies as a function of velocity and distance. By comparing the transfer function with different BLR models, the BLR geometry and kinematics can be inferred \citep[e.g.,][]{Bentz10,Grier13a,Xiao18a,Xiao18b,Horne20}. However, the interpretation could be nontrivial in some cases as different geometries can produce similar observational features.

Forward modeling of BLRs using phenomenological models is a powerful alternative approach \citep{Pancoast11,Brewer11}. \citet{Pancoast14a} developed the code {\small\texttt{CARAMEL}}  to implement this method, employing simple yet flexible parameterizations to describe the gas motion, and emissivity, as well as the gravitational influence of the BH. Through a diffusive nested sampling process, the model parameters and their uncertainties can be constrained, enabling a quantitative characterization of BLR geometry and kinematics.  More importantly, as an explicit parameter in the model, the BH mass can be directly constrained through the modeling. By comparing the BH masses with the virial products, one can directly calibrate the $f$ factor for each AGN without adopting a sample-averaged value or assuming that AGNs follow the same $M_{\rm BH}$--$\sigma_*$ relation as quiescent galaxies.

\begin{table*}[ht]
\centering
\caption{Sample Properties and Observation Parameters}  \label{tab:Observation}
    
    \begin{tabular}{r c c c c c c c}

 \hline \hline
Object name &  $z$ &  Monitoring Period &  $N_{\rm spec}$ &  $N_{\rm photo}$ &  $\Delta T_{\rm spec}$ &  $\Delta T_{\rm photo}$ &  SAMPID \\ 
(1) & (2) & (3) & (4) & (5) & (6) & (7) & (8) \\ 
 \hline
J0140$+$234 & 0.320 & 57353-59426 & 87 & 232 & 14 & 4 & Pr1\_ID03 \\
PG~0947$+$396 & 0.206 & 57355-59345 & 72 & 268 & 14 & 4 & Pr1\_ID15 \\
J1026$+$523 & 0.259 & 57412-59341 & 73 & 272 & 15 & 5 & Pr1\_ID17 \\
J1120$+$423 & 0.226 & 57383-59402 & 48 & 260 & 16 & 4 & Pr1\_ID23 \\
PG~1121$+$422 & 0.225 & 57383-59408 & 61 & 234 & 17 & 4 & Pr1\_ID24 \\
J1217$+$333 & 0.178 & 57415-59401 & 60 & 247 & 15 & 4 & Pr1\_ID29 \\
PG~1427$+$480 & 0.220 & 57426-59426 & 57 & 227 & 14 & 5 & Pr2\_ID24 \\
J1540$+$355 & 0.164 & 57426-59426 & 81 & 237 & 16 & 4 & Pr1\_ID41 \\
\hline
\multicolumn{8}{l}{\parbox{13cm}{Notes. Column (1): object names in the order of increasing R.A. Column (2): redshifts. Column (3): spectroscopic monitoring period. The number represents MJD. Column (4) and (5): $N_{\rm spec}$ and $N_{\rm photo}$ represents the total number of spectroscopic and photometric observations for each source, respectively. Column (6) and (7): the median cadence of the spectroscopic and photometric light curves, respectively.  Column (8): object internal ID assigned for each object as used in \cite{Woo19}. }}

\end{tabular}
\end{table*}

{\small\texttt{CARAMEL}}  modeling has been applied to $\sim 30$ AGNs from various RM campaigns, including the Lick AGN Monitoring Projects in 2008, 2011, and 2016 \citep{Pancoast14b, Williams18, Villafana22}, a 2010 AGN monitoring campaign at the MDM Observatory \citep{Grier17a}, and several individual sources \citep{Williams22, Bentz21, Bentz22, Bentz23b, Bentz23a}. Although {\small\texttt{CARAMEL}} has been primarily applied to \hbeta, it has also been used for higher-ionization emission lines such as \lya\ and \CIV\ \citep[e.g.,][]{Williams20}.
In addition to {\small\texttt{CARAMEL}}, \citet{Li13} developed an alternative implementation, {\small\texttt{BRAINS}}, which has been tested on many AGNs in parallel, yielding consistent results \citep{Li13, Li2018, Stone24}. 

Results from these studies indicate that BLRs in these AGNs are typically face-on, consistent with expectations for type 1 AGNs. Dynamically, BLRs can contain a mixture of gravitationally bound gas clouds and inflowing/outflowing components. More importantly, the virial factor derived from dynamical modeling is generally consistent with those inferred based on the $M_{\rm BH}$--$\sigma_*$ relation, as demonstrated in earlier dynamical modeling studies \citep{Pancoast14b, Grier17a, Williams18}. This agreement provides independent support for the commonly adopted assumption that active and inactive galaxies follow the same $M_{\rm BH}$--$\sigma_*$ relation.
However, these studies are based on relatively small samples and limited $M_{\rm BH}$ dynamic range, limiting the statistical strength of this conclusion and highlighting the need for larger datasets for confirmation.

The Seoul National University AGN Monitoring Project (SAMP) have monitored 32 moderate- to high-luminosity AGNs from 2015 to 2021 \citep{Woo19,Rakshit19,Cho23,Woo24,Wang25}. The final \hbeta\ lags of the 32 AGNs based on the 6-year monitoring are reported by \citet{Woo24}. \citet{Wang25} present the velocity-resolved lags for 20 SAMP AGNs with sufficient cadence.  In this paper, we present the dynamical modeling results for a subsample of them. Section 2 details the spectroscopic and photometric monitoring data, along with the spectral decomposition method used to isolate \hbeta\ from the AGN spectrum. Section 3 briefly outlines the geometrical and dynamical model of {\small\texttt{CARAMEL}} . Section 4 describes the modeling results for each AGN. In Section 5, we combine our sample with earlier dynamical modeling samples to constrain the mean virial factor. Finally, Section 6 provides a summary of our conclusions.

\section{Data}  \label{sec:observation}

\subsection{Photometric and Spectroscopic Data}

\begin{figure*}[ht]
\includegraphics[width=0.99\textwidth]{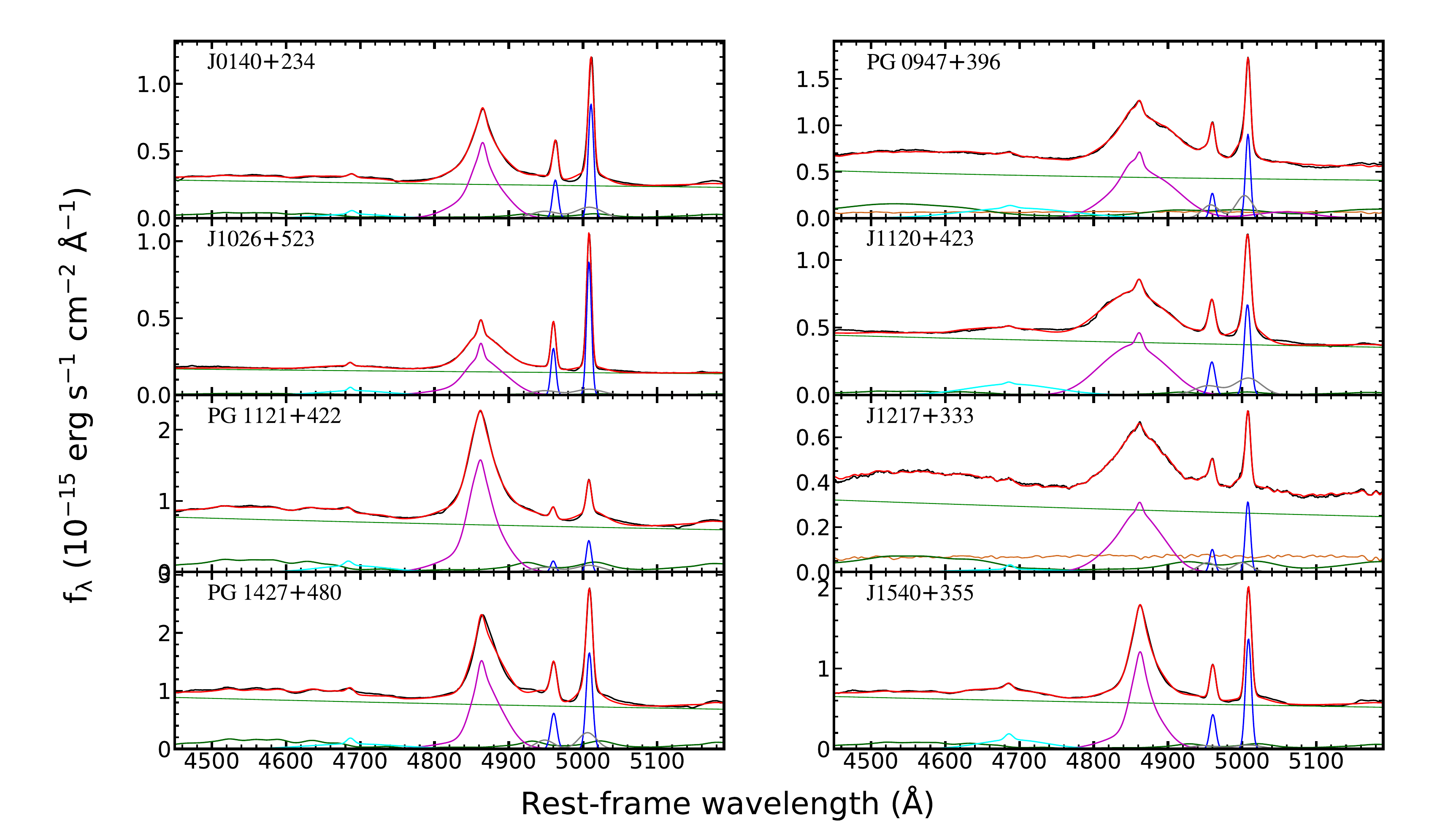}
\caption{Spectral decomposition of mean spectra for the eight AGNs modeled in this work. The continuum of each quasar is decomposed into a power law component (green), an Fe pseudocontinuum (dark green) based on the template provided by \citet{Boroson92}, and a host galaxy component (brown). The emission-line components consist of  \hbeta\ (magenta), narrow (blue) and wing  (gray) \OIII, and \HeII\ (cyan) emission lines. } \label{fig:decomposition}
\end{figure*}

Our initial target selection comprises 100 moderate- to high-luminosity AGNs drawn from the MIllIQUAS catalog \citep[Milliquas v4.5 (2015) update][]{Flesch2015} and the Palomar-Green (PG) quasar sample  \citep{Boroson92}. Following an initial variability assessment, we selected 32 AGNs as the final sample for continuous monitoring over a six-year period.  This sample spans a redshift range of $z=[0.079, 0.343]$ and a luminosity range of $L_{\rm 5100,AGN}=10^{44.1\sim45.6}$ erg s$^{-1}$. Detailed properties of these targets are summarized in \citet{Woo24}.  

Photometric observations were conducted using multiple facilities, including the MDM 1.3 m and 2.4 m telescopes, the 1 m class telescopes at Lemmonsan Optical Astronomy Observatory, Lick Observatory, Deokheung Optical Astronomy Observatory, as well as the Las Cumbres Observatory Global Telescope (LCOGT) network. Depending on the redshift of each target, observations were carried out using either B/V or  V/R filters \citep{Woo24}, in order to probe a similar rest-frame  wavelength range across all sources. Aperture photometry was performed using the {\small\texttt{SExtractor}} package \citep{BA96}, with apertures initially set to three times the seeing. The light curves from different telescopes were intercalibrated using the {\small\texttt{PyCALI}} code \citep{Li14}. For seven AGNs with relatively sparse sampling, we supplemented our continuum light curves with
$g$-band photometry \citep[see][for more details]{Mandal25} from the Zwicky Transient Facility \citep{Bellm19}.

Spectroscopic monitoring was conducted using the Lick Shane 3 m telescope and the MDM 2.4 m telescope. The Lick observations employed the Kast double spectrograph, which consists of separate arms for the blue and red wavelength ranges. In this study, only the red-side spectra were used to analyze the region around the \hbeta\ adjacent region. The red side observation employed a 600 lines mm$^{-1}$ grating and a  4$^{\prime\prime}$ slit to minimize the flux loss, yielding a spectral resolution of $R$=624. 
Due to a CCD upgrade in September 2016, the spectral coverage and dispersion of the Lick setup changed from 4300--7100\AA\ and 2.33 \AA/pixel before 2016 September to 4450--7280\AA\ and 2.17\AA/pixel afterwards. 
MDM observations used the Volume Phase Holographic blue grism, providing a wavelength coverage of 3970--6870\AA\ and a dispersion of 0.715\AA/pixel. 
We used a 3$^{\prime\prime}$ slit before 2017 January after which we ordered a customized 4$^{\prime\prime}$ slit in order to provide a consistent setup with Lick observation. The typical signal-to-noise ratio (S/N) of single-epoch spectra is was 15-20 per pixel.

The spectra were aligned in wavelength and flux by aligning their \OIII\ emission line profile to a reference spectrum \citep{VG92}, implemented via the {\tt mapspec} package \citep{Fausnaugh17}. After that, to account for site-dependent flux calibration differences between MDM and Lick observations, we applied minor corrections based on closely spaced observational pairs. The final absolute flux calibration was achieved by matching synthetic V band with the photometric V band light curves (R band for two high redshift AGNs) using {\small\texttt{PyCALI}} \citep{Li14}. 

Based on calibrated single-epoch spectrum, we constructed the mean and root-mean-square (rms) spectra. We performed multi-component decomposition to the mean spectra. The fit results of the eight sources included in this work are displayed in Figure \ref{fig:decomposition}. This sample consists of sources for which the dynamical modeling analysis yielded robust results (a more detailed description of the sample selection is provided in Section 4). The same decomposition procedure was applied to individual-epoch spectra to construct the broad \hbeta\ light curves.  We followed a modified version of the multi-component fitting technique developed by \citet{Guo18,Shen19}. See also the most recent update in \citet{Ren24}. The model pseudocontinuum consists of a power-law AGN continuum, an Fe II emission template, and a host galaxy starlight component. The continuum fitting is performed over two line-free wavelength windows: [4450, 4600] ${\rm \AA}$ and [5050, 5550] ${\rm \AA}$. We allow velocity shifts and Gaussian broadening for both the Fe II and stellar population components. Two commonly used Fe II templates from \citet{Boroson92} and \citet{Kovacevic10} were tested. Due to the use of multiple spectral settings of different wavelength coverage, we adopted the Boroson \& Green template as our default. The host galaxy component is modeled with a 10 Gyr, solar-metallicity single stellar population from \citet{BC03}. In most cases, host absorption features are negligible due to the relatively high AGN luminosities. For some sources with relatively lower luminosity (e.g., J1217+333), clear stellar absorption (such as the Mg Ib triplet) is detected and the host component is added.

After subtracting the best-fit pseudocontinuum, we modeled the emission-line residuals using multiple Gaussians. The broad \hbeta\ profile was fit with a sum of three Gaussians, while the narrow \hbeta\ component was modeled with a single Gaussian. The [O III] $\lambda$4959,5007
lines were fit with two Gaussians each to capture both core and wing components. We also included narrow and broad Gaussian components for \HeII\ with all narrow-line centroids and widths tied to one another (see Figure \ref{fig:decomposition}).

\begin{figure*}[htbp]
    \centering
    \includegraphics[width=0.9\textwidth]{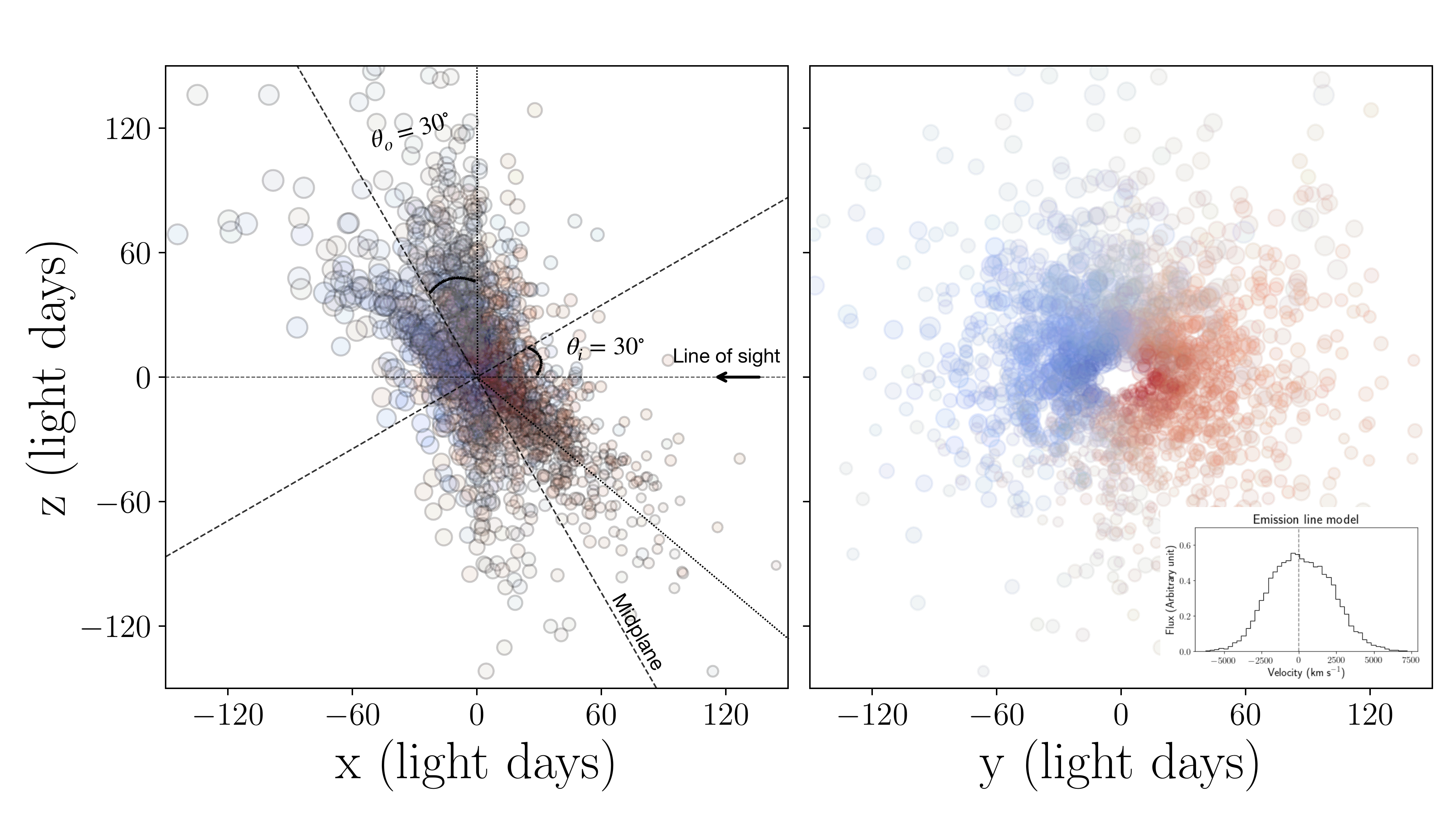}
    
    \caption{Illustration of the BLR model used in {\tt CARAMEL}.
Left panel: projection onto the $x$–$z$. The dashed arrow indicates the direction of the observer’s line of sight.  The BLR is viewed at an inclination angle of $\theta_i = 30^\circ$ and has an opening angle of $\theta_o = 30^\circ$, defined with respect to the BLR disk midplane. The symbol size represents the relative line emissivity, demonstrating a case where the emission is stronger on the far side of the BLR relative to the observer. The reduced number of clouds observed below the mid-plane results from the obscuration by the opaque mid-plane. 
Right panel: projection onto the $y$–$z$ plane, where the colors encode the line-of-sight velocity of each cloud, with blue and red indicating approaching and receding motions, respectively. The BLR consists of a combination of clouds on circular orbits and undergoing infalling motion. 
The inset shows the emission-line profile predicted by the model realization.}
    \label{fig:Example_Model}
\end{figure*}

\section{BLR Model} \label{sec:model}

\begin{table*}
\begin{minipage}{160mm}
 \caption{Summary of Key BLR Model Parameters and Their Interpretations}
 \label{table_params}
 \begin{tabular}{l l l}

  \hline
  Parameter & Definition & Interpretations \\
  \hline
  $r_{\rm mean}$ & \parbox[t]{7cm}{Mean value of the Mean radius $\mu$ posterior} &  \\
 $r_{\rm median}$ & \parbox[t]{8cm}{Median value of the Mean radius $\mu$ posterior}  &  \\
  $\beta$ & \parbox[t]{7cm}{Shape parameter of the radial profile} &  \parbox[t]{7cm}{$\beta=1.0$ indicates pure exponential profile; $\beta>1.0$, steeper; $\beta<1.0$, flatter.} \\
  $\theta_i$  & Inclination angle   & $\theta_i \rightarrow 0^{\circ}$: face on; $\theta_i \rightarrow 90^{\circ}$: edge on   \\
  $\theta_o$  & Opening angle  & $\theta_o \rightarrow 0^{\circ}$: razor-thin BLR; $\theta_o \rightarrow 90^{\circ}$: spherical BLR \\
  $\kappa$  & Cosine illumination function parameter   & \parbox[t]{8cm}{$\kappa < 0$: preferential emission from the farside of BLR; $\kappa > 0$: preferential emission from the nearside of BLR} \\
  $\gamma$  & Disk edge illumination parameter & \parbox[t]{8cm}{$\gamma \rightarrow 1$: emission uniformly distributed throughout the disk height; $\gamma \rightarrow 2$: emission clustered at the outer edges of BLR disk} \\
  $\xi$  & Plane transparency fraction & \parbox[t]{8cm}{$\xi \rightarrow0$: backside BLR fully obscured by the BLR midplane; $\xi \rightarrow 1$: no midplane obscuration} \\
  $M_{\rm BH}$  & Black hole mass &  \\
  $f_{\rm ellip}$  & Fraction of particles in circular-motion  orbits  &  \\
  $f_{\rm flow}$  &  \parbox[t]{7cm}{Flag indicating inflowing or outflowing orbits for the non-circular motion particles} & \parbox[t]{8cm}{$f_{\rm flow}<0.5$: the non-circular motion orbits particles are in inflowing orbits;  $f_{\rm flow}>0.5$: outflowing orbits}  \\
  $\theta_e$  & \parbox[t]{7cm}{Angle of the general velocity of the non-circular motion particles relative to the radial direction}   &  \parbox[t]{8cm}{$\theta_e \rightarrow 90^{\circ}$: nearly circular orbits; $\theta_e \rightarrow 0^{\circ}$: approaching escape velocity and are nearly unbound} \\
  In.$-$Out.   &  $\mathrm{sgn}(f_{\text{flow}} - 0.5) \times (1 - f_{\text{ellip}}) \times \cos(\theta_e)$
 & \parbox[t]{8cm}{ Inflow and outflow indicator. In.$-$Out.$\rightarrow1$ indicate radial outflow and $\rightarrow -1$ indicate pure radial inflow. }  \\
 $\sigma_{\rm turb}$  & Standard deviation of turbulent velocities &   \\
  \hline
 \end{tabular}
 
\end{minipage}
\end{table*}

{\small\texttt{CARAMEL}}  employs a parametric, phenomenological model to describe the geometry and kinematics of the BLR. Full details can be found in \citet{Pancoast14a}. Here we provide a brief summary  In this model, the \hbeta-emitting BLR is represented as an ensemble of point particles orbiting a central, point-like continuum source that emits isotropically. Each particle reprocesses the incident ionizing radiation after a time delay set by its radial distance from the continuum source. The reprocessing is assumed to occur instantaneously, and the observed wavelength of the reprocessed \hbeta\ emission is determined by Doppler shifts relative to the observer. A time series of model spectra is generated based on the input continuum light curve and the spatial and velocity distribution of the BLR particles. These distributions are described by a set of parameters, which are constrained by comparing the observed and modeled spectra. We summarize the key BLR parameters adopted in the model in the following subsections, and provide a schematic diagram in Figure \ref{fig:Example_Model} to help illustrate the model configuration. 

\subsection{Geometry}
The radial distribution ($r$) of BLR gas particles is modeled using a Gamma probability density function:
\begin{equation}
    p(r| \alpha,\theta) \propto r^{\alpha-1} {\rm exp}\left( -\frac{r}{\theta} \right) \label{eq:1}
\end{equation}
where $\alpha$ is the shape parameter and $\theta$ is the scale parameter. The distribution is shifted from the origin by the Schwarzschild radius, $R_s = 2GM/c^2$, plus a minimum radius $r_{\rm min}$. To aid physical interpretation, the following parameter transformations are introduced:
\begin{equation}
    \mu = r_{\rm min} + \alpha \theta \label{eq:2}
\end{equation}
\begin{equation}
    \beta = \frac{1}{\alpha} \label{eq:3}
\end{equation}
\begin{equation}
    F = \frac{r_{\rm min}}{r_{\rm min}+\alpha\theta} \label{eq:4}
\end{equation}
With this transformation, $\mu$ represents the mean radius of the BLR, $\beta$ is the transformed shape parameter, and $F$ denotes the ratio of the minimum radius $r_{\rm min}$ to the mean radius. The standard deviation of the shifted Gamma distribution is given by $\sigma_r = \mu \beta (1 - F)$. The outer boundary of the BLR is truncated at a radius $r_{\rm out}$.

The angular distribution of the BLR is described by two key parameters: the opening angle $\theta_o$ and the inclination angle $\theta_i$. The opening angle $\theta_o$ determines the vertical thickness of the BLR, with $\theta_o \approx 0^\circ$ corresponding to a razor-thin disk and $\theta_o \approx 90^\circ$ approaching a spherical configuration. The inclination angle $\theta_i$ is defined as the angle between the observer's line of sight and the BLR rotation axis. A value of $\theta_i = 0^\circ$ indicates a face-on view along the axis of rotation, while $\theta_i = 90^\circ$ corresponds to an edge-on view along the disk plane.

Three additional parameters are introduced to account for anisotropy in the \hbeta\ emission. First, the parameter $\gamma$ captures vertical asymmetry in the emissivity distribution. A value of $\gamma = 1$ corresponds to uniform emission throughout the vertical extent of the disk, while $\gamma = 2$ indicates that emission is preferentially concentrated in the outer layers. The parameter $\xi$ accounts for midplane obscuration, where $\xi \rightarrow 0$ implies complete obscuration of the far side of the BLR by the midplane, and $\xi = 1$ indicates no obscuration. Lastly, the parameter $\kappa$ describes the angular anisotropy of the reprocessed emission. A value of $\kappa = 0$ denotes isotropic emission, while $\kappa  <0$ and $\kappa > 0$ corresponds to preferential emission from far and near side as seen by the remote observer, respectively.

\subsection{Dynamics}
The velocity distribution of BLR particles is characterized by a combination of radial and tangential components. The parameter $f_{\rm ellip}$ denotes the fraction of particles on near-circular orbits that are gravitationally bound to the central black hole. The remaining fraction, $(1 - f_{\rm ellip})$, comprises particles following either inflowing ($f_{\rm flow} < 0.5$) or outflowing ($f_{\rm flow} > 0.5$) trajectories. Whether these motions remain bound or become unbound depends on the parameter $\theta_e$, which describes the orientation of the velocity vector within the plane defined by the radial and tangential velocity components. Specifically, $\theta_e$ quantifies the angular deviation between the particle velocity and the escape/circular velocity regimes. When $\theta_e = 0^\circ$, the velocity distribution peaks near the escape velocity, implying that most orbits are unbound. As $\theta_e$ approaches $90^\circ$, inflow and outflow velocities become increasingly tangential, approximating circular orbits. Intermediate values, such as $\theta_e \approx 45^\circ$, correspond to highly eccentric but still gravitationally bound trajectories, while lower values suggest that a majority of particles may be on unbound or marginally bound paths. Finally, we account for the contribution of macroturbulence in the BLR by adding an additional line-of-sight velocity component. The turbulent velocity is drawn from a normal distribution centered at zero with a standard deviation of $\sigma_{\rm turb}$, and then scaled by the absolute value of the particle’s circular velocity. The parameter $\sigma_{\rm turb}$ is dimensionless and defines the relative strength of turbulence velocity with respect to the circular velocity, which can range from
0.001 to 0.1.

For simplicity, we introduce a single parameter to characterize the BLR kinematics, referred to as In.$-$Out., which quantifies the degree to which the system exhibits radial inflow or outflow. It is defined as $\mathrm{sgn}(f_{\text{flow}} - 0.5) \times (1 - f_{\text{ellip}}) \times \cos(\theta_e)$, where a value of $+1$ corresponds to pure radial outflow and $-1$ corresponds to pure radial inflow.  

\subsection{Continuum Model and Implementation}

In the modeling, we represent the BLR with 2000 individual point particles. The observed continuum light curve is modeled using Gaussian processes, which allow for both interpolation between data points and extrapolation beyond the observation baseline. The interpolated continuum is then used to compute emission-line profiles at the epochs of actual spectroscopic observations in comparison with observation.

To compare the observed and modeled spectra, {\small\texttt{CARAMEL}}  employs a Gaussian likelihood function and determines the optimal BLR parameters using the diffusive nested sampling code {\small\texttt{DNEST4}} \citep{Brewer18}. We  incorporated a likelihood softening parameter, the statistical temperature $T$ to avoid any overfitting due to potential underestimation of flux uncertainties.  This is implemented by rescaling the log-scale likelihood by 1/$T$, which is equivalent to inflating the uncertainties by $\sqrt{T}$. Leveraging the properties of nested sampling, the effect of varying $T$ can be explored during post processing. In practice, the optimal $T$ is determined by inspecting parameter convergence plot where the posterior sample are plotted as a function of statistical levels. We identify the likelihood level where the sampling begins to exhibit ``streakiness"--a clear indicator of overfitting. We select a $T$ value that avoids this overfitted regime while maximizing the effective posterior sample size.

\section{Results} \label{sec:results}

From a sample of 32 AGNs that were continuously monitored for six years by SAMP, \citet{Woo24} identified a subset of 24 AGNs with reliable significant integrated H\(\beta\) lag detections based on a series of quantitative criteria. Among these, \citet{Wang25} selected 20 sources for velocity-resolved lag analysis, focusing those with a median spectroscopic cadence of \(< 20\) days.  In this study, we further refine the sample by selecting AGNs with a sufficient number of spectroscopic epochs (e.g., \(\gtrsim 50\)), considering the higher data quality requirements for dynamical modeling. As a result, {\small\texttt{CARAMEL}}  fitting was performed on 15 objects.

\begin{figure*}[htbp]
    \centering
    \includegraphics[width=0.49\textwidth]{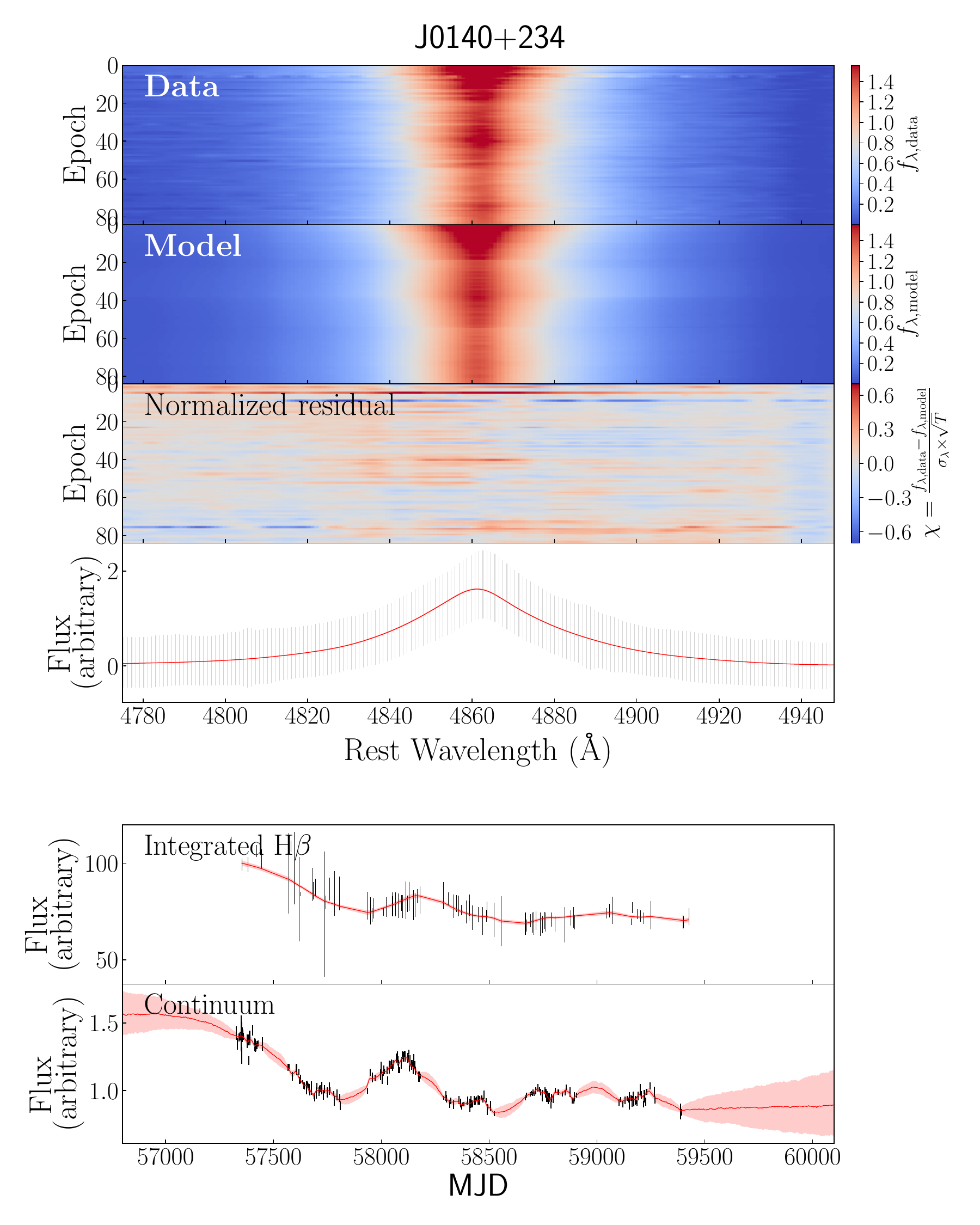}
    \includegraphics[width=0.49\textwidth]{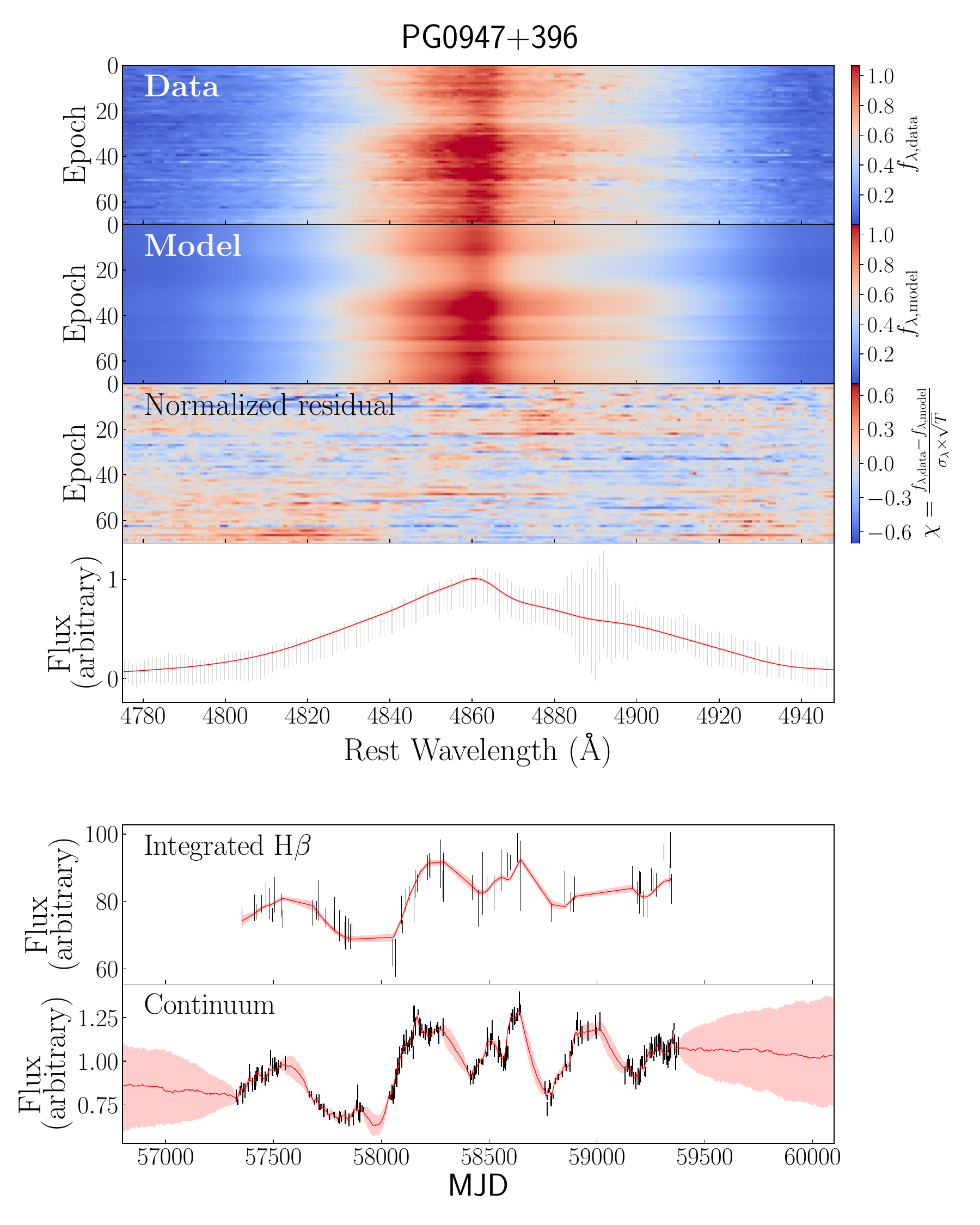}

    \includegraphics[width=0.49\textwidth]{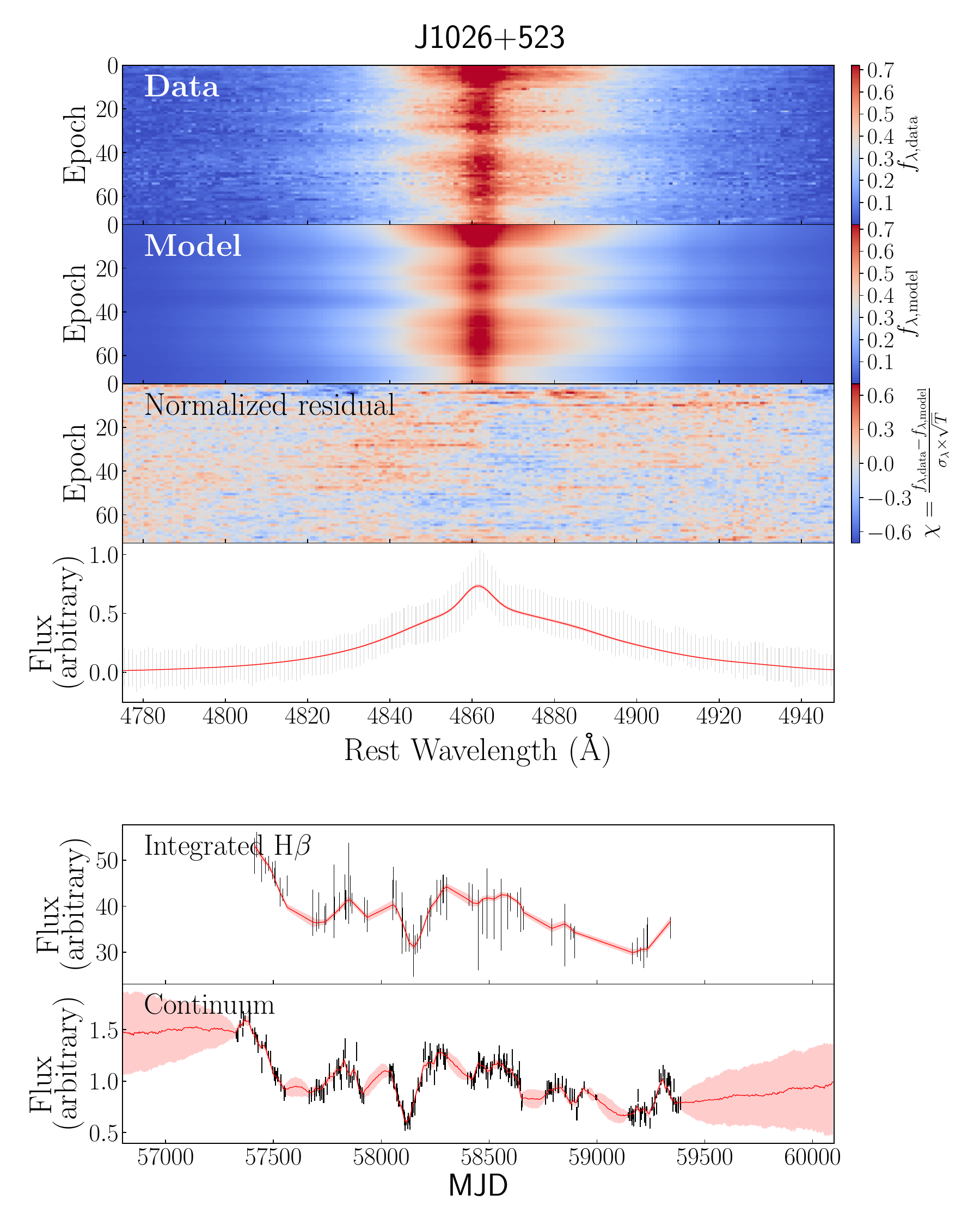}
    \includegraphics[width=0.49\textwidth]{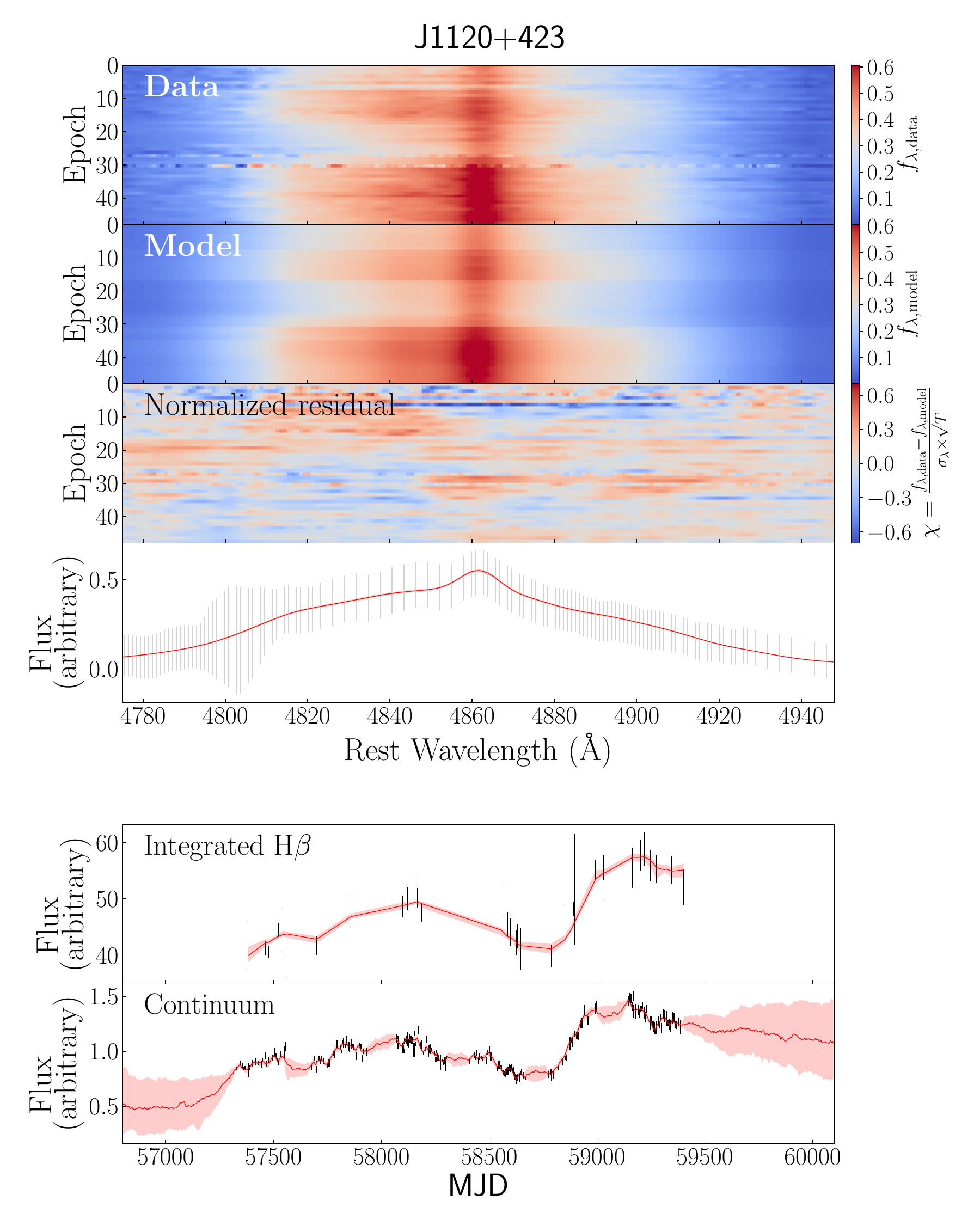}

    \caption{The {\small\texttt{CARAMEL}}  model fits to the \hbeta\ emission-line profile, integrated \hbeta\ light curves, and AGN continuum light curves. From top to bottom: Panels 1 and 2 show the observed and modeled \hbeta\ line profiles, ordered by observation epoch. Panel 3 presents the normalized residuals ($\chi$; see the main text for detailed description), while Panel 4 displays the observed \hbeta\ profile for a randomly selected epoch alongside the best-fit model (red). The error bars in this panel have been scaled by the statistical temperature $T$. Panels 5 and 6 show the  observed integrated \hbeta\ and AGN continuum light curves (black), along with the corresponding model fits (red). The results for  J0140+234, PG0947+396, J1026+523, and J1120+423 are shown.}
    \label{fig:Example_Fitting}
\end{figure*}

\addtocounter{figure}{-1}
\begin{figure*}[htbp]
    \centering

     \includegraphics[width=0.49\textwidth]{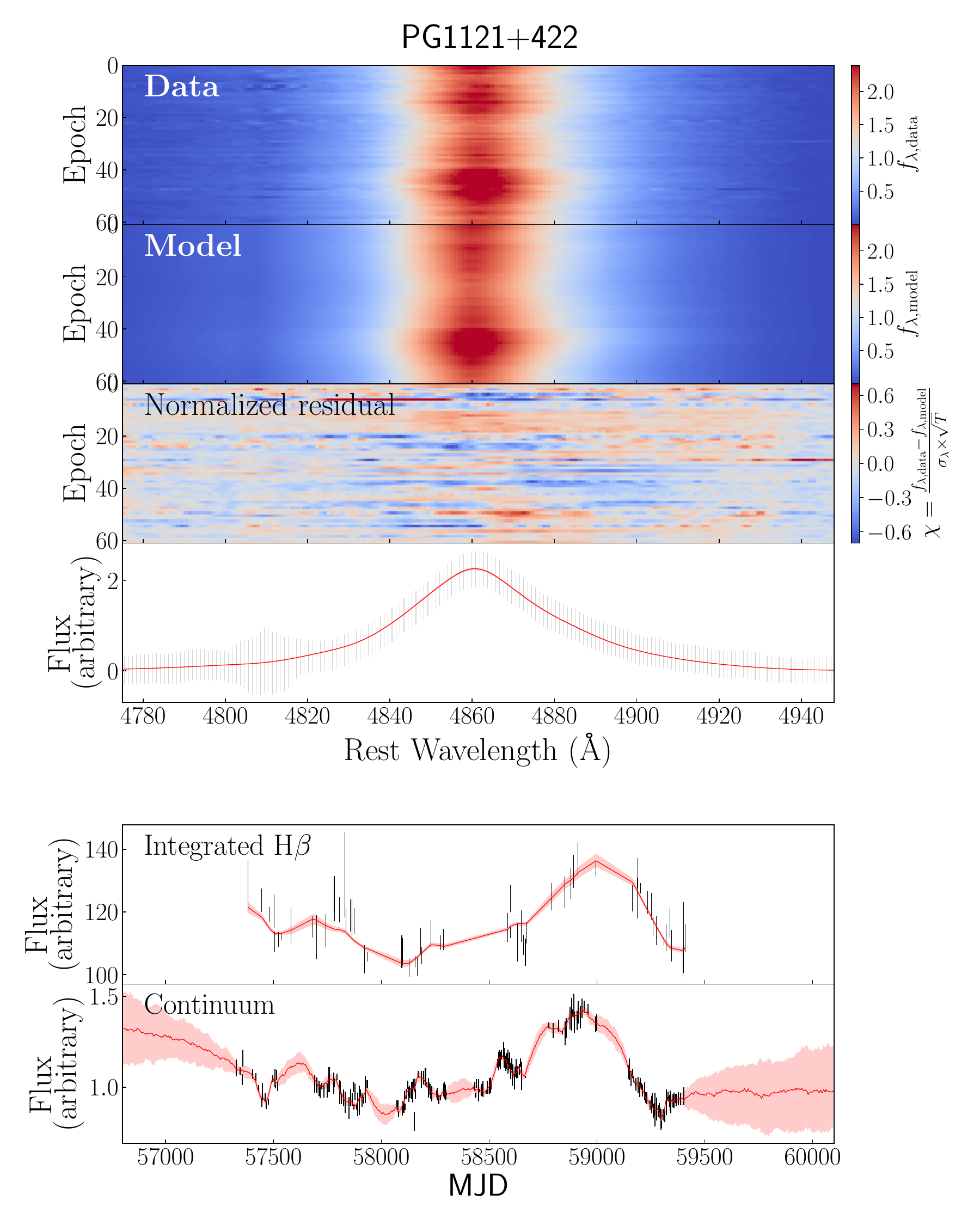}
    \includegraphics[width=0.49\textwidth]{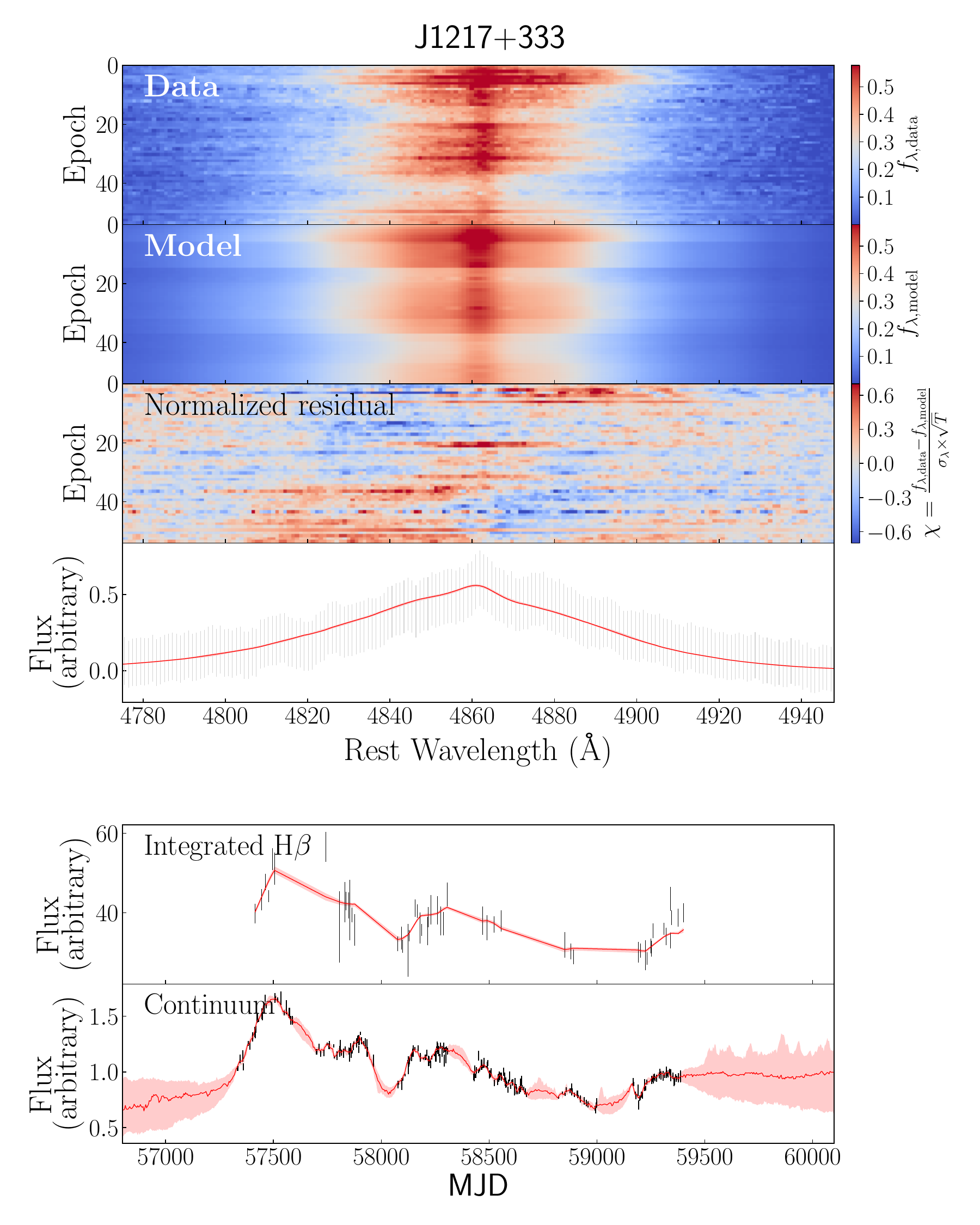}
    \includegraphics[width=0.49\textwidth]{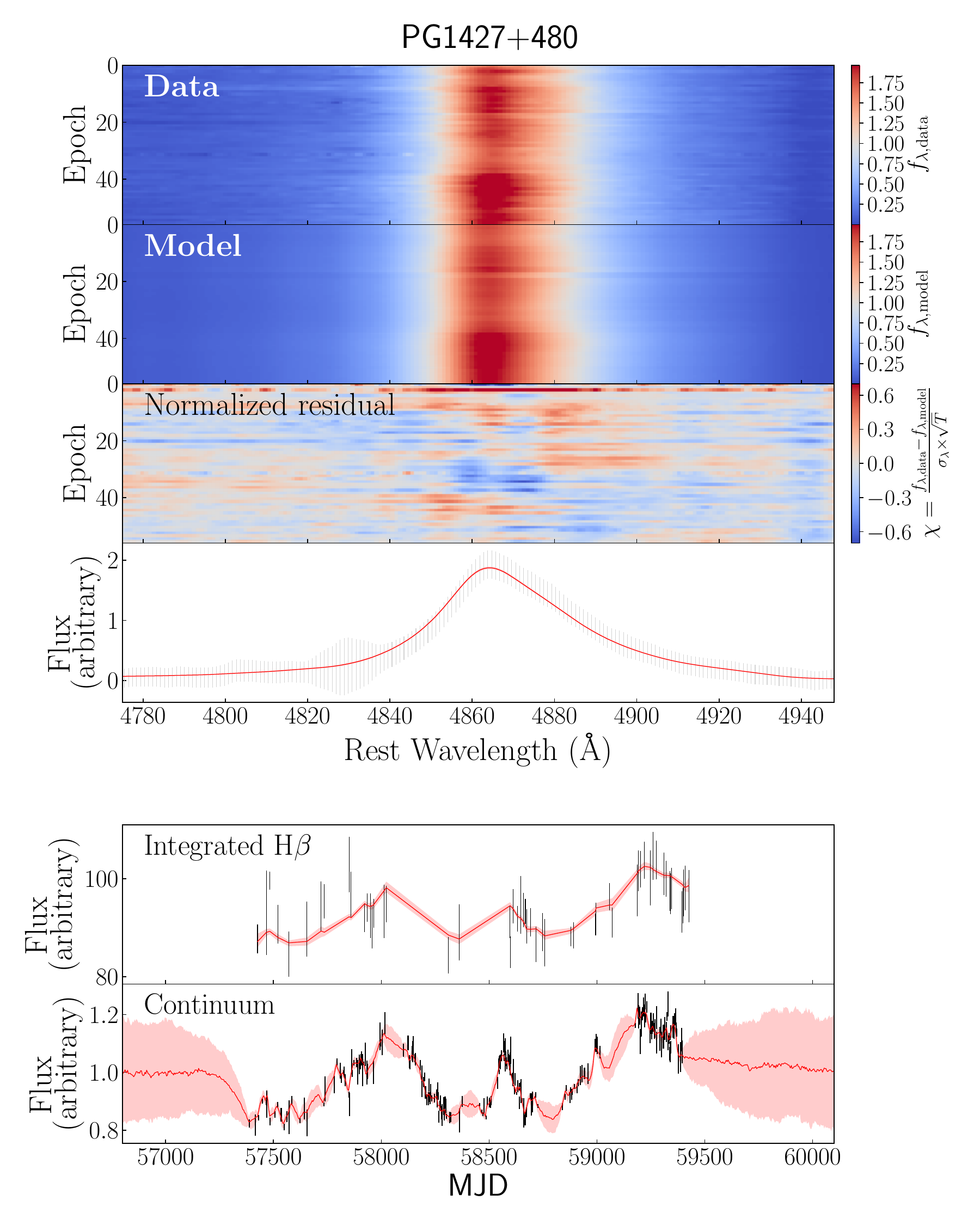}
    \includegraphics[width=0.49\textwidth]{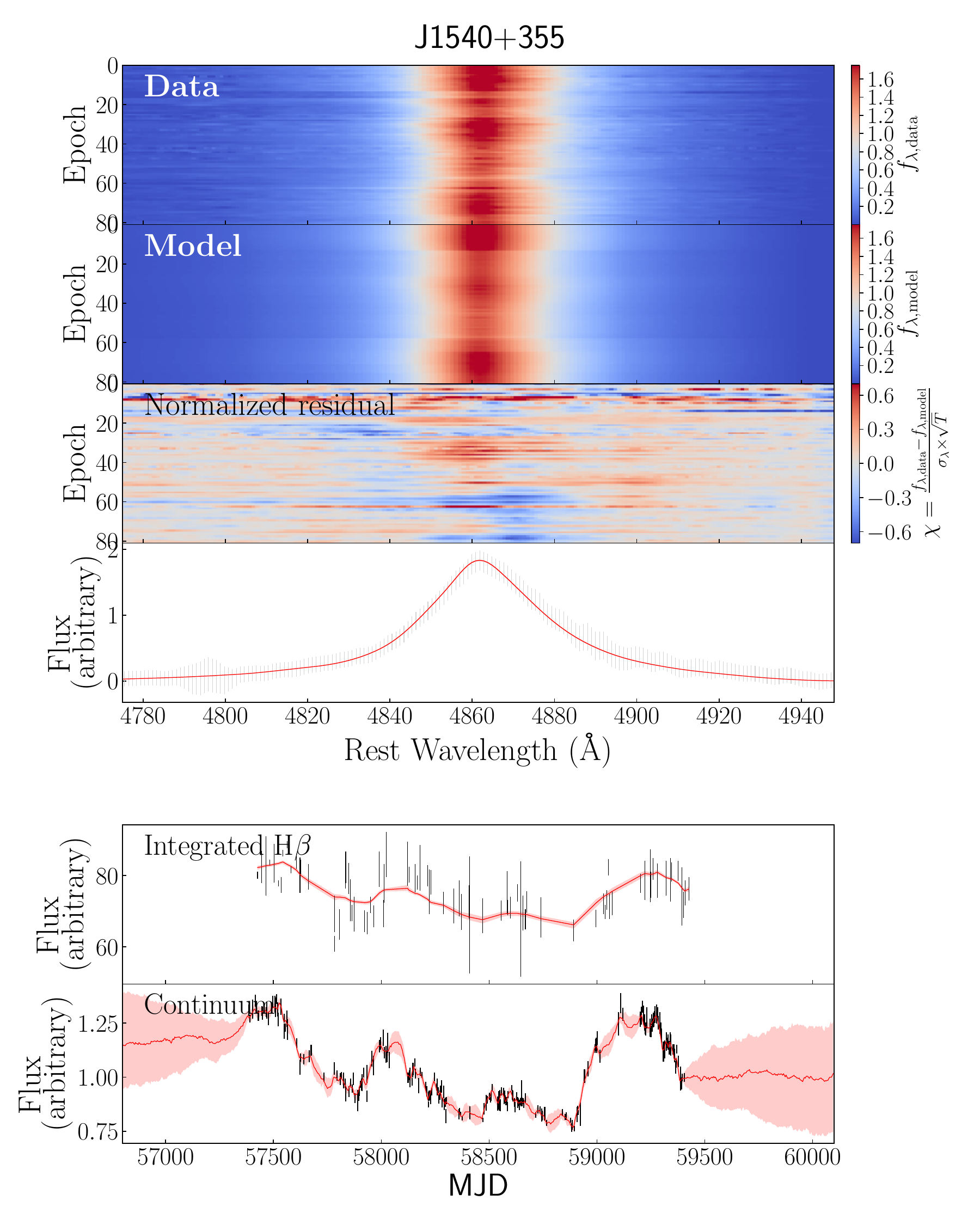}

    \caption{Continued. The results for PG 1121+422, J1217+333, PG1427+480, and J1540+355.} \label{fig:Example_Fitting2}
\end{figure*}

\begin{figure*}[htbp]
    \centering
    \includegraphics[width=0.95\textwidth]{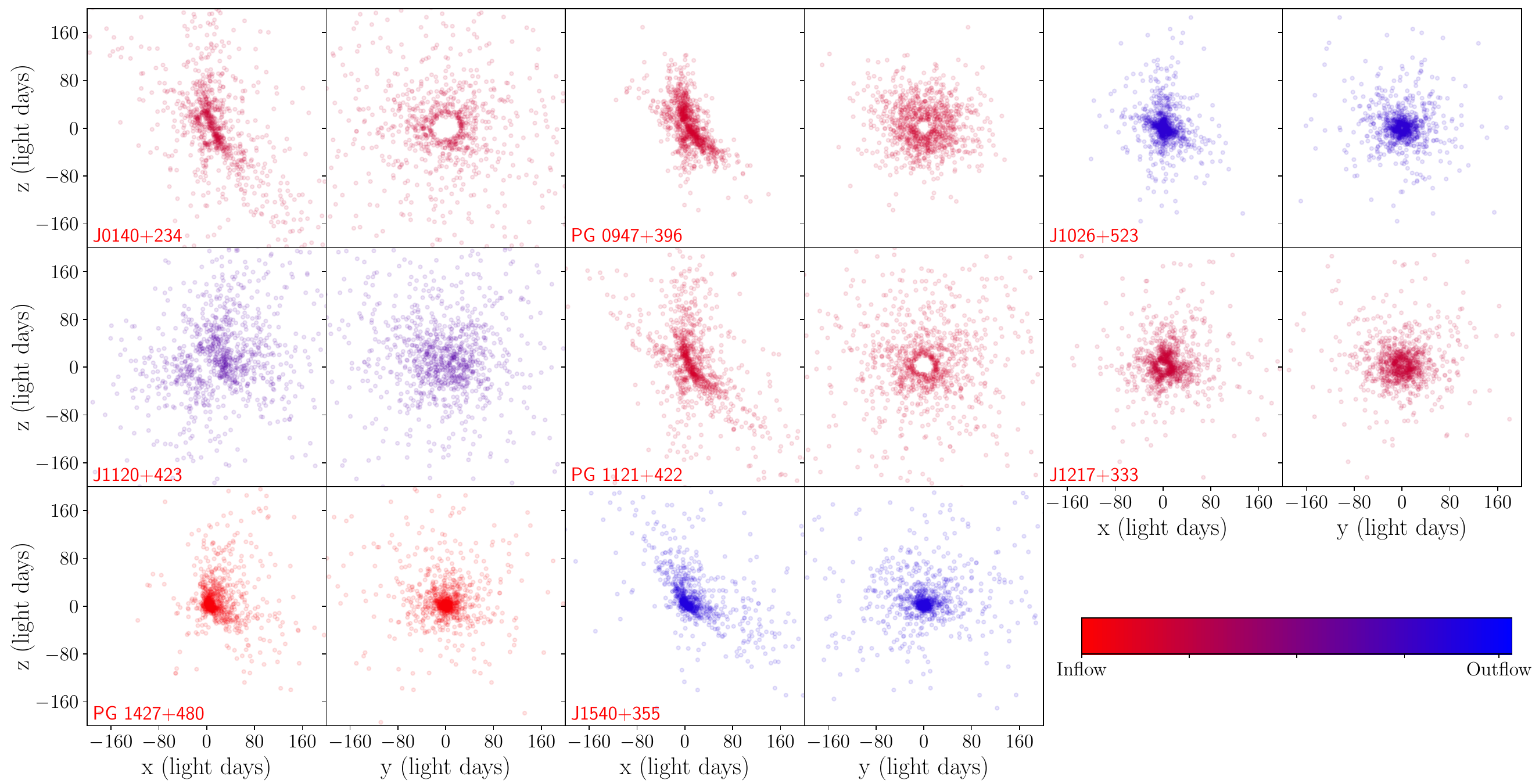}

    \caption{Geometric representation of BLR emission for the eight AGNs, based on median parameter estimates from {\small\texttt{CARAMEL}} . The left panel depicts an edge-on perspective for each source, while the right panel shows a face-on view. Each point in the plot represents a point-like gas cloud, with colors indicating their dynamics: red for inflow and blue for outflow.
}
    \label{fig:Geometry}
\end{figure*}

\begin{figure*}[htbp]
    \centering
    
    \includegraphics[width=1.0\textwidth]{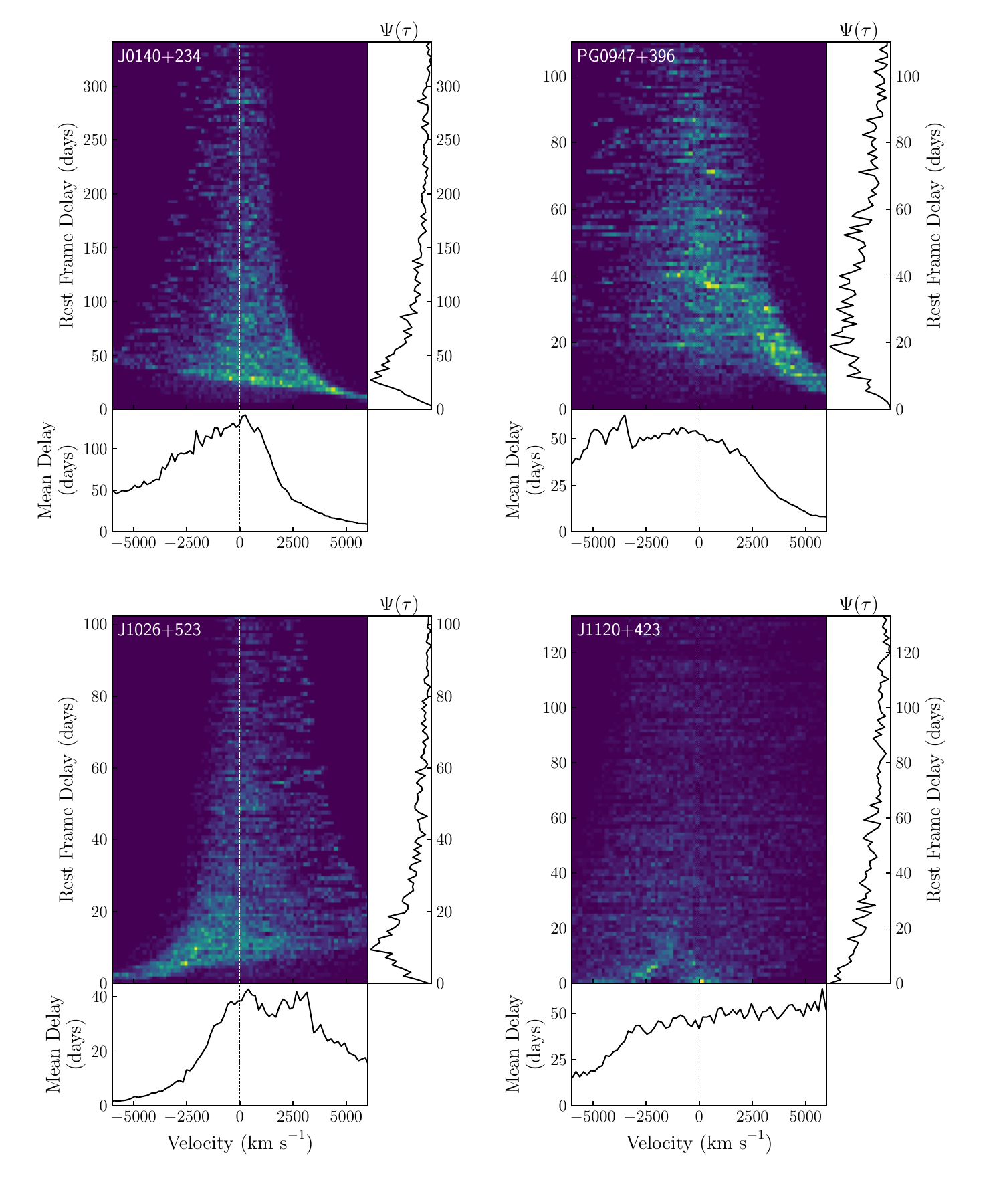}

    \caption{Transfer function  produced using median model parameter estimates. The right-hand panel shows the velocity-integrated transfer function and the bottom panel shows the average time lag for each velocity pixel. The results for  J0140+234, PG0947+396, J1026+523, and J1120+423 are shown. }
    \label{fig:Example_TF}
\end{figure*}
\addtocounter{figure}{-1}
\begin{figure*}[htbp]
    \centering
    
    \includegraphics[width=1.0\textwidth]{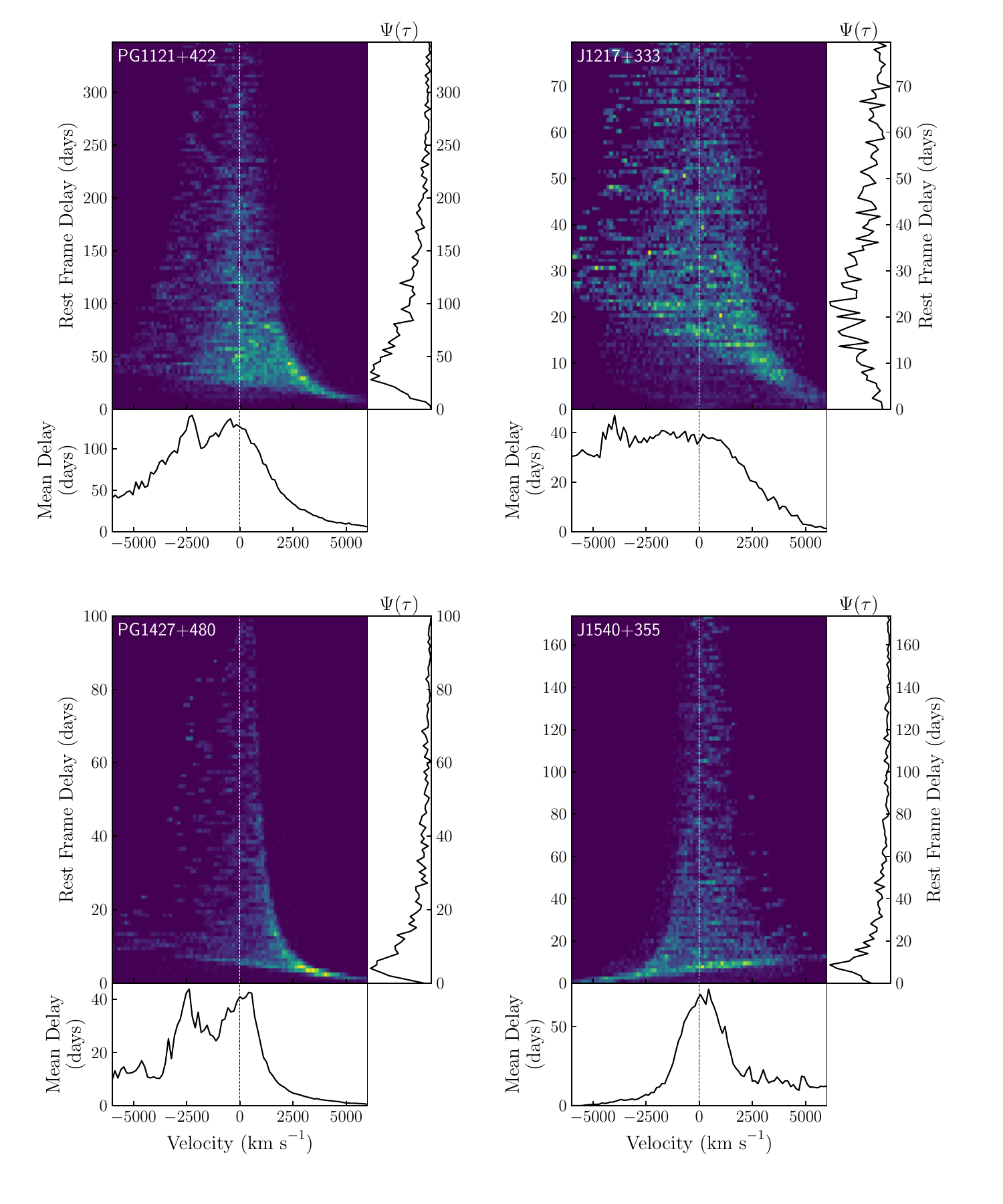}

    \caption{Continued. The results for PG 1121+422, J1217+333, PG1427+480, and J1540+355.}
\end{figure*}

By inspecting the fitting results and the posterior distribution, we exclude seven objects where either the {\small\texttt{CARAMEL}}  models exhibit only moderate fitting quality or the fitting is not converged. The fitting results for these excluded objects are provided and discussed in the Appendix \ref{sec:AppendixA}. 

In Figure \ref{fig:Example_Fitting}, we present the {\small\texttt{CARAMEL}}  fitting results for the eight objects that we categorize as good model fitting results. They are selected by a combination of inspection of the agreement between model fits and the integrated H\(\beta\) light curves, the 2-dimensional residuals (defined as $(f_{\rm \lambda, data}-f_{\rm \lambda, model})/(\sigma_{\rm \lambda}\times\sqrt{T})$),  as well as the convergence of the posteriors. The best-fit model parameters and their associated uncertainties are summarized in Table \ref{tab:model_parameters_good}.

Geometrically, the H\(\beta\)-emitting BLRs in our sample is best described as a thick disk with a typical opening angle of \(\sim\) 40 degrees, while three objects exhibit larger angles approaching 60 degrees (Figure \ref{fig:Geometry}). The BLRs  are viewed at intermediate inclinations, with typical inclination angles of about $30$ degrees.  The H\(\beta\) emission predominantly originates from the far side of the BLR, in agreement with previous dynamical modeling results.  The BLRs often show a transparent or slight obscuration in the midplane.  

Dynamically, the BLR gas exhibits a combination of circular, inflowing, and outflowing motions (Figure \ref{fig:Example_TF}). In many cases, significant contributions from inflow and outflow components are detected with the fraction of particles in elliptical orbits less than 50\%. The contribution from turbulence is generally negligible. In the following sections, we provide a detailed description of BLR properties on an object-by-object basis.

\subsection{J0140+234}

We calculated the reduced $\chi^2$ and the fractional root-mean-square difference (FRMSD) between the model and observed \hbeta\ light curves. For this object, we obtained a reduced $\chi^2$ of 0.93 and an FRMSD of 5\%, indicating that the model provides an excellent fit to the data. Geometrically, the radial profile is slightly steeper than a exponential profile with a shape parameter $\beta=1.31^{+0.45}_{-0.29}$ for the Gamma distribution. The median and mean radius are derived as $r_{\rm median}=91^{+24}_{-23}$ and $135^{+18}_{-12}$  light-days, respectively, with cross-correlation derived $R_{\rm BLR}$ of 113.6$^{+9.5}_{-10}$ light-days lying in the middle. The BLR is a moderately thick disk with an opening angle of $\theta_o = 27.2^{+11}_{-7.8}$ degrees, viewed at an inclination angle of $\theta_i = 21.7^{+9.8}_{-8.8}$ degrees relative to the observer's line of sight. The model suggests that the emission is preferentially emitted at the edge of the disk ($\gamma = 1.76^{+0.17}_{-0.25}$) and possibly from the far side of the BLR ($\kappa = -0.43^{+0.70}_{-0.06}$) although the uncertainty is huge. A partial opaque midplane ($\xi = 0.49=^{+0.27}_{-0.22}$) is implied.  

Dynamically, the results find two possible solutions for the fraction of particles in circular orbits ($f_{\rm ellip})$. The higher-probability solution suggests 39\% of the particles are in the circular orbits, while the lower-probability  solution suggests $\sim$75\% particles. Although we see double peak profile in $f_{\rm ellip}$,  the result generally implies  inflow characteristics for the remaining particles ($f_{\rm flow} = 0.34^{+0.40}_{-0.23}$) with the general velocity deviating slightly from radial direction ($\theta_e = 9.7^{+10}_{-6.6}$ degrees). Turbulent velocity contributions are found to be negligible ($\sigma_{\rm turb} = 0.02^{+0.05}_{-0.02}$). 

Due to the double-peak $f_{\rm ellip}$ posterior, the black hole mass also has two possible solutions, so the reported value exhibits a large upper uncertainty. Nevertheless, the derived log$_{10}(M_{\rm BH}/M_{\odot})=8.37^{+0.72}_{-0.14}$ is consistent with the value from traditional RM, i.e., log$_{10}(M_{\rm BH}/M_{\odot})=8.30^{+0.06}_{-0.06}$,  where $\sigma_{\rm line,rms}$ and the corresponding log$_{10}\,f=0.62\pm0.15$ is used \citep{Woo15}. 

\subsection{PG 0947+396}

We obtained a reduced $\chi^{2}$ of 0.99 and an FRMSD of 4\%, indicating that the model fits the data very well for this object. The radial profile of \hbeta\ emission in this object is flatter than an exponential distribution ($\beta=0.76^{+0.24}_{-0.17}$). The radius  is measured to be $r_{\rm median} = 47.5^{+6.0}_{-6.2}$ light-days and   $r_{\rm mean}=55.7^{+7.3}_{-7.0}$ light-days, with former closer to the value measured from cross-correlation analysis (36.7$^{+9.5}_{-11}$ light-days).  The BLR is characterized as a moderately thick disk ($\theta_o = 29.5^{+9.1}_{-7.5}$ degrees) and is viewed at relatively face-on inclination ($\theta_i = 28.1^{+8.5}_{-6.5}$ degrees). The model cannot constrain whether the \hbeta\ emission is edge-concentrated or not (as indicated by $\gamma = 1.59^{+0.29}_{-0.39}$), but we find a mild enhancement of the \hbeta\ emission from the far side of the disk ($\kappa = -0.26^{+0.13}_{-0.15}$), along with a partial transparent midplane ($\xi = 0.50^{+0.15}_{-0.12}$).

The dynamical model suggests that 28\% of the BLR particles are in near-circular orbits ($f_{\rm ellip} = 0.28^{+0.10}_{-0.13}$), while inflowing motion dominates among the remainder ($f_{\rm flow} = 0.26^{+0.17}_{-0.17}$). The general velocity of these inflowing particles deviates from radial direction by an angle of $\theta_e = 24^{+14}_{-12}$ degrees. The contribution from turbulent motion is minimal, with $\sigma_{\rm turb} = 0.01^{+0.03}_{-0.01}$.

The black holes mass is determined to be log$_{10}(M_{\rm BH}/M_{\odot})=8.26^{+0.21}_{-0.16}$, which is fully consistent with the traditional RM mass of log$_{10}(M_{\rm BH}/M_{\odot})=8.30^{+0.12}_{-0.16}$. 

\subsection{J1026+523}

We obtained a reduced $\chi^{2}$ of 0.32 and an FRMSD of 5\%. Although the small reduced $\chi^{2}_{\nu}$ might suggest a possible presence of overfitting, our visual inspection indicates that the model accurately captures the general variability features over the six-year span.  Furthermore, the low FRMSD of 5\%  confirms that the model is in high agreement with the data. The radial profile is slightly steeper than a exponential profile ($\beta=1.16^{+0.23}_{-0.20}$). The \hbeta\ emission region exhibits a median radius of $r_{\rm median} = 27.6^{+5.9}_{-5.0}$ light-days and a mean radius of $40.7^{+9.1}_{-6.1}$ light-days, with the cross-correlation $R_{\rm BLR}$ sitting in the middle (36.7$^{+9.5}_{-11}$ light-days).  The BLR takes the form of a thick disk, characterized by an opening angle of $\theta_o = 44.1^{+11.9}_{-14.4}$ degrees. The viewing angle is moderately inclined, at $\theta_i = 29.7^{+9.1}_{-12}$ degrees. There is no preference of emission in the disk-height direction ($\gamma = 1.49^{+0.34}_{-0.33}$), but there is a mild preferential of emission from the far side of the disk ($\kappa = -0.32^{+0.08}_{-0.12}$). The midplane is largely transparent, with $\xi = 0.83^{+0.12}_{-0.20}$.

The kinematic configuration reveals that 29\% of the test particles follow near-circular trajectories ($f_{\rm ellip} = 0.29^{+0.16}_{-0.16}$). Outflow motion dominates the remaining particles ($f_{\rm flow} = 0.74^{+0.18}_{-0.18}$), whose velocity deviating from the radial direction by an angle of $\theta_e = 24^{+19}_{-15}$ degrees. Turbulent velocities are found to be negligible, with $\sigma_{\rm turb} = 0.01^{+0.04}_{-0.01}$.

The black holes mass is determined to be log$_{10}(M_{\rm BH}/M_{\odot})=7.72^{+0.28}_{-0.16}$, which is  consistent with the traditional RM mass of log$_{10}(M_{\rm BH}/M_{\odot})=7.60^{+0.11}_{-0.11}$. 

\subsection{J1120+423}

The models fits the data reasonably well, yielding a reduced $\chi^{2}$ of 1.12 and an FRMSD of 4\%. The \hbeta\ emission region exhibits a median radius of $r_{\rm median} = 81^{+36}_{-43}$ light-days and a mean radius of $109^{+35}_{-63}$, which are both larger than the cross-correlation values (44$^{+17}_{-15}$ light-days).  The BLR in this source is inferred to be close to sphere, with an opening angle of $\theta_o = 67.8^{+13}_{-6.5}$ degrees. The inclination angle toward the observer is moderately high, measured at $\theta_i = 40.8^{+14}_{-8.4}$ degrees. light-days. There is a preference of emission coming from the edge of the disk ($\gamma = 1.66^{+0.28}_{-0.38}$). A slight near-side enhancement in line emission is evident ($\kappa = 0.16^{+0.22}_{-0.28}$), while the midplane shows partial transparency ($\xi = 0.59^{+0.17}_{-0.32}$).

In terms of dynamics, 51\% of the test particles exhibit near-circular motion ($f_{\rm ellip} = 0.51^{+0.17}_{-0.37}$). The remainder are dominated by outflowing trajectories ($f_{\rm flow} = 0.67^{+0.22}_{-0.17}$) but may still remain in gravitationally bound orbits  as indicated by $\theta_e = 48^{+23}_{-30}$ degrees. The contribution from turbulent motion is negligible, with $\sigma_{\rm turb} = 0.01^{+0.03}_{-0.01}$.

The black holes mass is determined to be log$_{10}(M_{\rm BH}/M_{\odot})=8.36^{+0.35}_{-0.25}$, which is fully consistent with the traditional RM mass (log$_{10}(M_{\rm BH}/M_{\odot})=8.26^{+0.17}_{-0.17}$).

\subsection{PG 1121+422}

The model fits the data very well, yielding a reduced $\chi^{2}$ of 1.22 and an FRMSD of 4\%.  According to the model, the radial profile is close to exponential ($\beta=1.05^{+0.20}_{-0.17}$). The \hbeta\ emission region exhibits a median radius of $r_{\rm median} = 83.5^{+8.0}_{-5.8}$ light-days and a mean radius of $111.4^{+13}_{-6.6}$ light-days, with the latter closer to the cross-correlation results (116$^{+24}_{-20}$ light-days).  The BLR in this source is shaped as a moderately thick disk, defined by an opening angle of $\theta_o = 29.4^{+7.5}_{-9.0}$ degrees and observed at an inclination of $\theta_i = 24.7^{+8.3}_{-8.0}$ degrees. The emission is slightly concentrated from the edge of the disk ($\gamma = 1.77^{+0.17}_{-0.34}$). There is modest preferential emission from the far side of the disk ($\kappa = -0.43^{+0.20}_{-0.06}$). The midplane transparency is not well constrained.

The dynamical model suggests that 33\% of the gas follows stable circular orbits ($f_{\rm ellip} = 0.36^{+0.09}_{-0.10}$), while the remaining component shows inflowing motion ($f_{\rm flow} = 0.26^{+0.15}_{-0.18}$) with a slight angular offset from radial direction of $\theta_e = 10.7^{+7.8}_{-7.5}$ degrees. Turbulence appears minimal, with $\sigma_{\rm turb} = 0.06^{+0.03}_{-0.05}$. 

The black holes mass is determined to be log$_{10}(M_{\rm BH}/M_{\odot})=8.18^{+0.29}_{-0.15}$, which is fully consistent with the traditional RM mass of log$_{10}(M_{\rm BH}/M_{\odot})=8.04^{+0.12}_{-0.08}$. 

\subsection{J1217+333}

The model fits the data reasonably, yielding a reduced $\chi^{2}$ of 1.53 and an FRMSD of 8\%. Its radial profile is slightly steeper than exponential. The \hbeta\ emission region exhibits a median radius of $r_{\rm median} = 38.5^{+4.9}_{-6.6}$ light-days and a mean radius of $55.9^{+4.6}_{-6.9}$ light-days, with the former closer to the cross-correlation values (26.5$^{+21.2}_{-20.7}$ light-days). The geometry of the BLR is best described by a very thick disk with an opening angle of $\theta_o = 63.4^{+22.4}_{-16.6}$ degrees. It is viewed at an inclination of $\theta_i = 31.6^{+6.1}_{-10.0}$ degrees.  The vertical profile is relatively uniform, characterized by $\gamma = 1.36^{+0.39}_{-0.24}$. Emission appears to be significantly enhanced on the far side of the disk ($\kappa = -0.44^{+0.14}_{-0.03}$), while the midplane is partially transparent, with $\xi = 0.66^{+0.26}_{-0.16}$.

Kinematically, the model reveals that 40\% of the BLR material move on circular orbits ($f_{\rm ellip} = 0.40^{+0.25}_{-0.29}$). The rest is dominated by inflowing trajectories, with $f_{\rm flow} = 0.18^{+0.23}_{-0.12}$ and a modest angular offset from radial direction, $\theta_e = 14^{+10}_{-9}$ degrees. The turbulence component is negligible, with $\sigma_{\rm turb} = 0.02^{+0.05}_{-0.02}$.

The black holes mass is determined to be log$_{10}(M_{\rm BH}/M_{\odot})=7.82^{+0.24}_{-0.11}$, which is fully consistent with the traditional RM mass of log$_{10}(M_{\rm BH}/M_{\odot})=7.84^{+0.31}_{-0.31}$. 

\subsection{PG 1427+480}

The model fits the data generally well, yielding a reduced $\chi^{2}$ of 1.34 and an FRMSD of 4\%. It shows a steeper radial profile than the exponential function. The \hbeta\ emission region exhibits a median radius of $r_{\rm median} = 25^{+11}_{-13}$ light-days and a mean radius of $46^{+25}_{-21}$ light-days, where the cross-correlation value sits in the middle (34$^{+21}_{-19}$ light-days). This source exhibits a very thick BLR disk, with an opening angle of $\theta_o = 57^{+15}_{-17}$ degrees and is viewed at an inclination of $\theta_i = 25.3^{+19}_{-9.4}$ degrees. The emission doesn't show a preference of being uniform or edge-concentrated in the vertical direction ($\gamma = 1.56^{+0.35}_{-0.45}$), but it is strongly enhanced from the far side of the disk ($\kappa = -0.49^{+0.05}_{-0.01}$), where the midplane shows a strong opacity, with $\xi = 0.17^{+0.08}_{-0.06}$.

On the dynamical side, only a small fraction of the gas—about 5\%—follows circular orbits ($f_{\rm ellip} = 0.05^{+0.06}_{-0.04}$). The rest gas predominantly exhibits inflow behavior ($f_{\rm flow} = 0.31^{+0.30}_{-0.19}$) with slight angular offset from radial direct by $\theta_e = 7.9^{+4.8}_{-5.2}$ degrees. Turbulence plays a negligible role, with $\sigma_{\rm turb} = 0.01^{+0.03}_{-0.01}$.

The black holes mass is determined to be log$_{10}(M_{\rm BH}/M_{\odot})=7.46^{+0.24}_{-0.28}$, which is smaller but still consistent within $1 \sigma$ uncertainty with the traditional RM mass (log$_{10}(M_{\rm BH}/M_{\odot})=7.70^{+0.26}_{-0.26}$). 

\subsection{J1540+355}

The model fits obtains a reduced $\chi^{2}$ of 2.82 and an FRMSD of 6\%.  Although reduced $\chi^{2}$ is relatively large due to the fail of modeling the second-year variability,  visual inspection confirms that the general variability features over the six-year period are well captured. Consequently, we categorize this object as a good fit for our subsequent analysis. The \hbeta\ emission region shows a steeper than exponential radial profile ($\beta=1.58^{+0.16}_{-0.14}$).  It exhibits a mean radius of $r_{\rm median} = 80^{+29}_{-22}$ light-days, and a median radius of $39^{+12}_{-13}$ light-days, where the cross-correlation value is in the middle (58$^{+18}_{-15}$ light-days).  The BLR in this source is characterized by a relatively thick disk, with an opening angle of $\theta_o = 47.2^{+14}_{-8.6}$ degrees and viewed at a relatively large inclination of $\theta_i = 45.1^{+7.5}_{-18}$ degrees.  The vertical distribution of emission is relatively uniform, as indicated by $\gamma = 1.22^{+0.22}_{-0.16}$. Emission is preferentially enhanced on the far side of the BLR ($\kappa = -0.45^{+0.09}_{-0.04}$), and the midplane shows relatively high opacity of $\xi = 0.17^{+0.22}_{-0.13}$.

Kinematically, about 24\% of the BLR gas follows circular orbits ($f_{\rm ellip} = 0.24^{+0.23}_{-0.14}$). The rest shows clear outflow signatures ($f_{\rm flow} = 0.66^{+0.22}_{-0.28}$), with the velocity vector deviating modestly from radial direction by $\theta_e = 14.0^{+11}_{-8.9}$ degrees. The contribution from turbulent motion is negligible, with $\sigma_{\rm turb} = 0.02^{+0.05}_{-0.01}$.

The black holes mass is determined to be log$_{10}(M_{\rm BH}/M_{\odot})=7.46^{+0.24}_{-0.16}$, which is consistent within $1 \sigma$ uncertainty with the traditional  reverberation mass of log$_{10}(M_{\rm BH}/M_{\odot})=7.78^{+0.15}_{-0.15}$.



\section{Discussion} \label{sec:discussion}

\subsection{Constraining the virial factor}
\begin{table*}
\centering
\caption{Logarithmic Virial Factors Based On Four Kinds of Line Widths for Individual AGNs in the SAMP Sample.}
\label{tab:individual_f}
 \begin{tabular}{lrrrr}
\hline \hline
Object &  \multicolumn{4}{c}{log$_{10}\,f$} \\ \cline{2-5}
 & $\rm \sigma_{\rm line, rms}$ & ${\rm FWHM_{rms}}$ & ${\rm \sigma_{\rm line,mean}}$ & ${\rm FWHM_{\rm mean}}$\\  \hline
J0140$+$234 & $0.74_{-0.21}^{+0.40}$ & $0.35_{-0.19}^{+0.41}$ & $0.63_{-0.18}^{+0.42}$ & $0.12_{-0.19}^{+0.42}$ \\
PG~0947$+$396 & $0.70_{-0.23}^{+0.23}$ & $0.04_{-0.25}^{+0.26}$ & $0.65_{-0.22}^{+0.23}$ & $-0.03_{-0.22}^{+0.23}$ \\
J1026$+$523 & $0.75_{-0.20}^{+0.22}$ & $0.07_{-0.21}^{+0.25}$ & $0.33_{-0.20}^{+0.23}$ & $-0.26_{-0.20}^{+0.23}$ \\
J1120$+$423 & $0.85_{-0.30}^{+0.26}$ & $0.10_{-0.30}^{+0.27}$ & $0.74_{-0.30}^{+0.26}$ & $-0.06_{-0.30}^{+0.26}$ \\
PG~1121$+$422 & $0.83_{-0.20}^{+0.22}$ & $0.16_{-0.2}^{+0.22}$ & $0.43_{-0.20}^{+0.22}$ & $0.02_{-0.19}^{+0.22}$ \\
J1217$+$333 & $0.62_{-0.27}^{+0.51}$ & $0.07_{-0.30}^{+0.51}$ & $0.48_{-0.27}^{+0.52}$ & $-0.22_{-0.28}^{+0.51}$ \\
J1540$+$355 & $0.37_{-0.25}^{+0.29}$ & $-0.19_{-0.24}^{+0.28}$ & $0.09_{-0.24}^{+0.28}$ & $-0.32_{-0.24}^{+0.28}$ \\
PG~1427$+$480 & $0.40_{-0.47}^{+0.40}$ & $0.11_{-0.45}^{+0.39}$ & $0.22_{-0.44}^{+0.39}$ & $-0.26_{-0.45}^{+0.39}$ \\
\hline
\multicolumn{5}{l}{\parbox{0.51\textwidth}{Note. The first column lists the object names. The values in the following four columns represent the logarithm virial factors based on four different line width definitions indicated by the column headers. These values were obtained by comparing the $M_{\rm BH}$ posterior samples from {\small\texttt{CARAMEL}}  with the virial products derived from lag and line width measurements from \citet{Woo24}. }}
\end{tabular}
\end{table*}

One of the main objectives of this study is to constrain the virial factor $f$ using the $M_{\rm BH}$ directly determined from dynamical modeling. For each AGN, $f$ can be calculated by comparing the modeled $M_{\rm BH}$ with the virial product derived from the  time lag and line width measured from RM. This enables an independent calibration of the sample-averaged $f$ for AGNs that can be compared with the traditional one based on $M_{\rm BH}$--$\sigma_*$ relation.

The virial factor can have different values based on different line width definitions. There are four commonly used line width definitions:  $\sigma_{\rm line}$ and FWHM measured from the rms spectrum ($\sigma_{\rm line, rms}$, FWHM$_{\rm rms}$) or the mean spectrum ($\sigma_{\rm line, mean}$, FWHM$_{\rm mean}$). Among these, $\sigma_{\rm line, rms}$ is generally regarded as the most robust line width for RM $M_{\rm BH}$ estimates \citep[e.g.,][]{Peterson2004, Collin06, Wang19,Dalla-Bonta20}, and is the most widely adopted in RM studies. However, for single-epoch $M_{\rm BH}$ estimates, rms spectra are not available, so the virial factors based on $\sigma_{\rm line, mean}$ or FWHM$_{\rm mean}$ are used instead.

The lag and line width measurements of SAMP AGNs are provided in Table 6 of \citet{Woo24}. In this work, we calculate the four types of virial products and determine the corresponding virial factors. To estimate the uncertainties in $f$ from the full $M_{\rm BH}$ posteriors, we generate a set of mock lags and line widths by drawing random samples from Gaussian distributions centered on the reported measurements, with standard deviations set by their uncertainties. For asymmetric uncertainties—such as those in lag measurements—we adopt the average of the upper and lower errors as the standard deviation. The number of mock realizations is matched to the size of the $M_{\rm BH}$ posterior sample, enabling a direct derivation of the posterior distribution of $f$. The final $\langle f \rangle$ is taken as the median of the derived $f$ distribution, and the lower and upper uncertainties are defined by the 16th to 50th and 50th to 84th percentile intervals, respectively. The resulting virial factors for individual SAMP AGNs are summarized in Table~\ref{tab:individual_f}. 

With the virial factor distributions derived for a sample, we proceed to determine the sample-averaged $f$ and its intrinsic dispersion. 
We adopt a hierarchical Bayesian approach following \citet[][see the appendix for details]{Pancoast14b}.
In brief, we assume that the observed virial factors is a sample drawn from an intrinsic $f$ distribution that is a Gaussian distribution characterized by a mean $\langle \log_{10} f \rangle$ and a dispersion of $\sigma_{\log_{10} f}$. 
By evaluating how well different combinations of $\langle \log_{10} f \rangle$ and $\sigma_{\log_{10} f}$ reproduce the observed virial factor distribution for our sample, we can derive the two-dimensional posterior distribution of these two parameters, from which their best estimates and uncertainties are obtained.

In addition,  we compute a predictive distribution of virial factors, denoted as log$_{10}(f)_{\rm pred}$, which incorporates both the uncertainty in the mean virial factor and intrinsic dispersion of the $f$ distribution, i.e., the object by object variation. 

The log$_{10}(f)_{\rm pred}$  represents the expected probability distribution from which future virial factors are drawn, given our current knowledge.

\begin{figure*}[htbp]
    \includegraphics[width=0.99\textwidth]{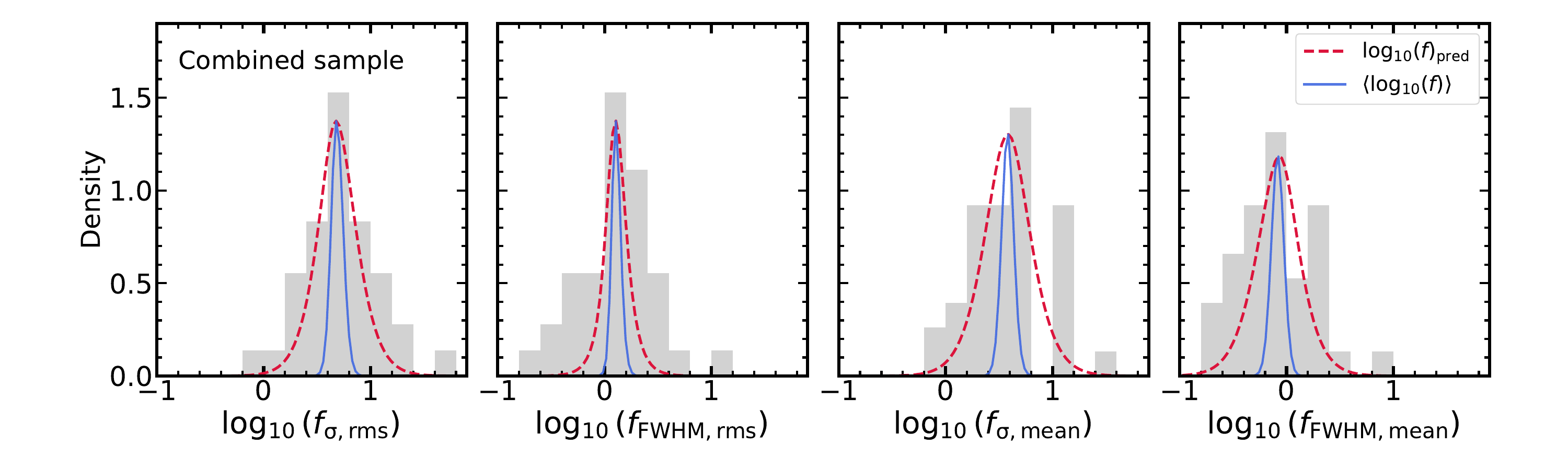}
    \caption{The predictive distribution of virial factors (red dashed) along with the posterior distribution of the mean virial factors (blue solid). The gray histograms represent the distribution of virial factors  calculated from individual sources. 
    The results presented here are based on the 38 AGNs from the combined sample, and the four panels represent the results for different types of line widths. 
} \label{fig:f_distribution}
\end{figure*}

In addition to the SAMP sample, 30 AGNs with {\small\texttt{CARAMEL}}  modeling are available in the literature \citep[noting that NGC 5548 has two independent measurements]
{Pancoast14b, Grier17a, Williams18, Williams20, Bentz21, Bentz22, Bentz23a, Bentz23b}.  A large portion of this sample has been summarized in detail by \citet[][]{Villafana23}, with two additional AGNs reported in \citet{Bentz23a} and \citet{Bentz23b}. Combining the eight AGNs from SAMP with these literature cases yields a total of 38 AGNs with dynamical modeling measurements consistently derived using {\small\texttt{CARAMEL}} .

For this combined sample, we collected the {\small\texttt{CARAMEL}} -derived posterior samples of $M_{\rm BH}$ for each object through private communication, and derived the  log$_{10}(f)_{\rm pred}$ using the method described above. With the inclusion of SAMP sample, this sample is not only  larger in size but also extends to higher-$M_{\rm BH}$ regime, enabling a more representative description for overall AGN population.
The derived virial factors for the combined sample are summarized in Table \ref{tab:meanf_SAMP}, and Figure \ref{fig:f_distribution} shows the predictive distribution of $f$ together with the posterior distribution of the mean $f$ estimates.   
The results based on the SAMP sample and the previous literature sample is also calculated for completeness. 

The log$_{10}(f)_{\rm pred}$ based on $\sigma_{\rm line, rms}$, $\sigma_{\rm line, mean}$, FWHM$_{\rm rms}$ and FWHM$_{\rm mean}$ is $0.69\pm0.21$, $0.58\pm0.25$, $0.11\pm0.14$ and  $-0.08\pm0.23$, respectively, based on combined sample. When converted to linear scale,  these $f$ values are $4.90\pm2.37$, $3.80\pm2.18$, $1.29\pm0.42$, and $0.83\pm0.44$. 
With the inclusion of SAMP sample, the average virial factors remain consistent with previous values, but the dispersion is slightly decreased. 
As a consistency check, we also perform a linear regression between the dynamical modeling and the canonical RM masses using \texttt{Linmix} \citep{Kelly07} (see Figure 7).  This yields a consistent virial factor of 0.67$\pm$0.08 and a slightly larger but still consistent intrinsic scatter of 0.27$\pm$0.09 dex. We adopt the dispersion of the Bayesian analysis (e.g., 0.21 dex for $\sigma_{\rm line,rms}$) as the representative intrinsic scatter, as this approach fully incorporates the information from the posterior distributions of the dynamical masses. However,   in the case where the assumed Gaussian form cannot describe the observed virial factor distribution (e.g., for FWHM$_{\rm rms}$), the virial factor should be interpreted as a lower limit.
Lastly,  we note that virial factors from the rms-spectra line widths are larger than those of corresponding line width from mean spectra, which can be understood because the rms line widths are generally $\sim$20\% smaller than the mean spectra line widths \citep[e.g.,][]{Collin06,Barth15}. 

These dynamical modeling-based virial factors are in agreement with previous calibrations based on the $M_{\rm BH}$--$\sigma_*$ relation. Using a sample of 16 RM AGNs, \citet{Onken04} reported an average value of $\langle f \rangle = 5.5 \pm 1.9$ based on $\sigma_{\rm line}$. \citet{Woo10} later updated the calibration using 24 AGNs, deriving log$_{10}f = 0.72^{+0.09}_{-0.10}$ for $\sigma_{\rm line}$ with an intrinsic scatter of $0.44 \pm 0.07$ dex. More recently, \citet{Woo15} revised these values using an updated RM AGN sample, reporting log$_{10}f = 0.65 \pm 0.12$ for $\sigma_{\rm line}$ and log$_{10}f = 0.05 \pm 0.12$ for FWHM.

These previous calibrations rely on the assumption that active and inactive galaxies follow the same $M_{\rm BH}$–$\sigma_*$ relation. The virial factors in this work are derived independently from dynamical modeling; therefore the consistency between these two approaches supports the validity of this assumption.

\begin{figure}[htbp]
    
    \includegraphics[width=0.5\textwidth]{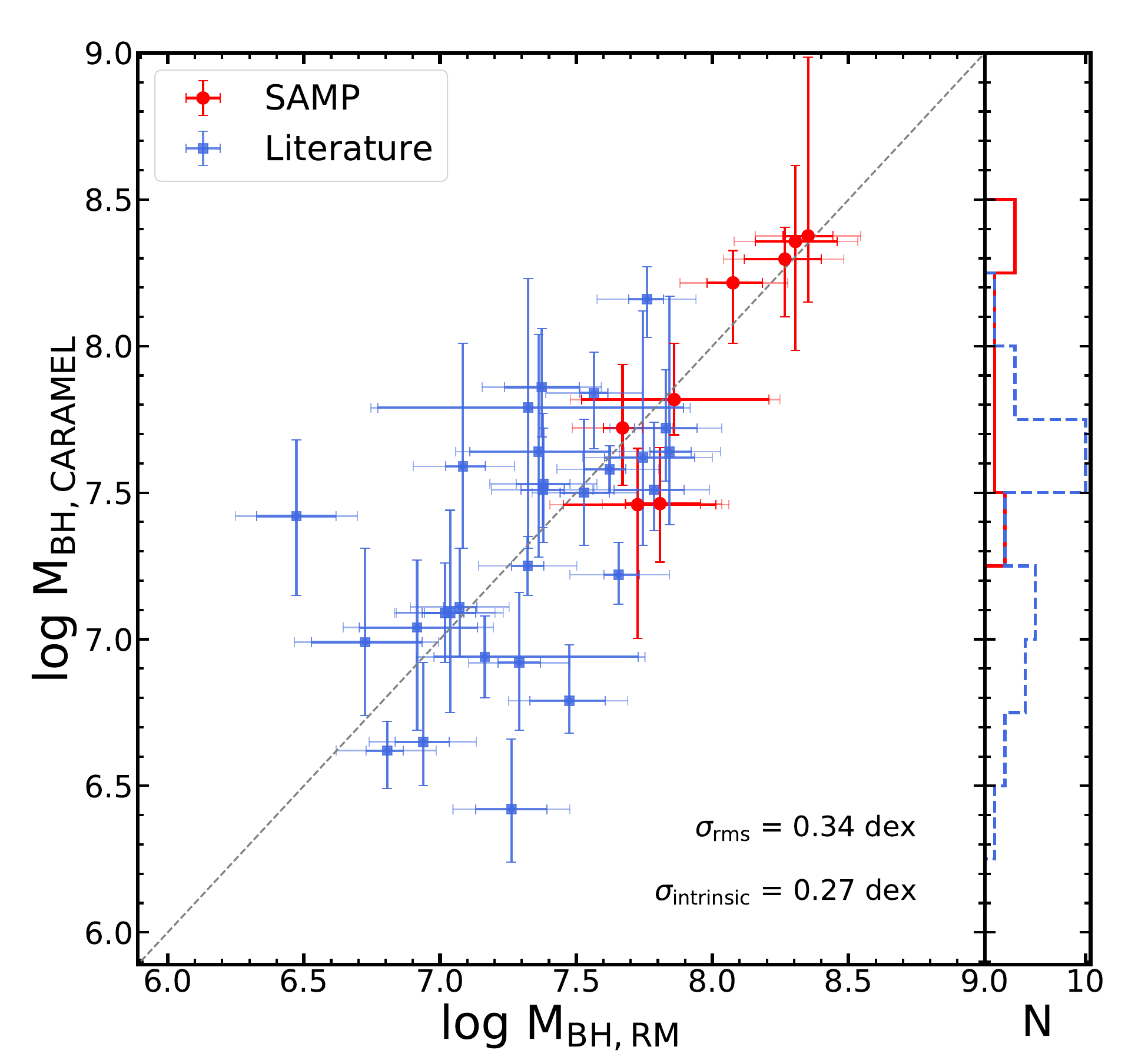}
    \caption{Comparison between black hole masses derived from {\small\texttt{CARAMEL}}  modeling and traditional RM-based estimates. Red and blue symbols denote the SAMP and literature samples, respectively. The RM-based masses are calculated by applying the updated virial factor from this work (log$_{10}(f)_{\rm pred} = 0.69\pm0.21$) to the virial products. Two sets of error bars are shown: solid bars represent statistical uncertainties, while dashed bars include the additional systematic uncertainty associated with the virial factor. The histogram on the right panel shows the mass distributions of the SAMP and literature samples. The root-mean-square (rms) and intrinsic scatter of the mass comparison are indicated in the lower right corner.} 
    \label{fig:Mass_comparison}
\end{figure}

It has been reported that the apparent $M_{\rm BH}$–$\sigma_*$ relation for AGNs appears shallower than that of quiescent galaxies, which is caused by selection biases that limit
the $M_{\rm BH}$ dynamic range at $ > 10^8,M_{\odot}$ \citep{Woo13, Winkel25}. Our sample mostly consists of relatively high-mass AGNs, which extends the $M_{\rm BH}$ dynamic range at the $M_{\rm BH}\sim 10^8$ to $10^{8.5} M_{\odot}$ range (Figure~\ref{fig:Mass_comparison}). The fact that no significant difference in the virial factor is found between the SAMP and literature samples suggests that $f$ remains consistent even in this high-mass regime, where the constraints from the AGN $M_{\rm BH}$–$\sigma_*$ relation remain ambiguous. 

Notably, the intrinsic scatter of virial factors estimated through dynamical modeling (e.g., the dispersion in $\log_{10} (f)_{\rm pred}$ of 0.21 dex for $\sigma_{\rm line,rms}$) is smaller than that of the AGN $M_{\rm BH}$–$\sigma_*$ relation \citep[0.43 dex;][]{Woo10,Woo15,Winkel25}. This could imply that calibrating virial factors using the AGN $M_{\rm BH}$–$\sigma_*$ relation likely results in overestimated  systematic uncertainty for RM $M_{\rm BH}$. By adopting the updated $f$ factors,  the systematic uncertainty in RM $M_{\rm BH}$ is refined to 0.2 $\sim$ 0.25 dex.  while for the single-epoch $M_{\rm BH}$ the systematic uncertainty would be approximately 0.3 dex after including the $\sim$0.23 dex intrinsic scatter of the BLR size--luminosity relation 
\citep[e.g.,][]{Wang24,Woo24}.


\begin{table*}[htbp]
\centering
\caption{Summary of $\langle {\rm log}_{10}(f)\rangle$, $\sigma_{{\rm log}_{10}\,f}$, and log$_{10}(f)_{\rm pred}$ based on the SAMP and the Combined Sample.}
\label{tab:meanf_SAMP}
 \begin{tabular}{lllrrrr}
\hline \hline
Sample &  $N_{\rm object}$ &  & $\sigma_{\rm line, rms}$ & FWHM$_{\rm rms}$ & $\sigma_{\rm line, mean}$ & FWHM$_{\rm mean}$ \\ \hline 
\multirow{3}{*}{Literature} & \multirow{3}{*}{30} &  $\langle$log$_{10}(f)\rangle$ & $0.71\pm0.07$ & $0.14\pm0.06$ & $0.62\pm0.07$ & $-0.06\pm0.07$ \\
&  & $\sigma_{{\rm log}_{10}(f)}$ & $0.26\pm0.08$ & $0.15\pm0.07$ & $0.27\pm0.08$ & $0.23\pm0.08$ \\
 &  & log$_{10}(f)_{\rm pred}$ & $0.71\pm0.28$ & $0.14\pm0.17$ & $0.62\pm0.29$ & $-0.06\pm0.25$ \\  \hline
 
\multirow{3}{*}{SAMP} & \multirow{3}{*}{8}  &  $\langle$log$_{10}(f)\rangle$ & $0.67\pm0.12$ & $0.11\pm0.11$ & $0.45\pm0.11$ & $-0.11\pm0.11$ \\
& &  $\sigma_{{\rm log}_{10}(f)}$ & $0.17\pm0.11$ & $0.16\pm0.10$ & $0.18\pm0.13$ & $0.17\pm0.11$ \\
 & &  log$_{10}(f)_{\rm pred}$ & $0.67\pm0.23$ & $0.11\pm0.22$ & $0.45\pm0.25$ & $-0.11\pm0.23$ \\  \hline
 \multirow{3}{*}{Combined} & \multirow{3}{*}{38} &  $\langle$log$_{10}(f) \rangle$ & $0.69\pm0.06$ & $0.12\pm0.04$ & $0.58\pm0.06$ & $-0.08\pm0.06$ \\
& &  $\sigma_{{\rm log}_{10}(f)}$ & $0.20\pm0.06$ & $0.12\pm0.05$ & $0.23\pm0.06$ & $0.21\pm0.06$ \\
 & & log$_{10}(f)_{\rm pred}$ & $0.69\pm0.21$ & $0.11\pm0.14$ & $0.58\pm0.25$ & $-0.08\pm0.23$ \\
\hline
\multicolumn{7}{l}{\parbox{0.7\textwidth}{Note. Summary of the virial factor distribution based on the literature, SAMP, and the combined sample. $N_{\rm object}$ denotes the number of objects used in the calculation. $\langle {\rm log}_{10}(f)\rangle$ and $\sigma_{{\rm log}_{10}\,f}$ represent the mean and the dispersion of the Gaussian function of the intrinsic virial factor distribution, and log$_{10}(f)_{\rm pred}$ describes the predictive distribution for future BH mass estimation. }}
\end{tabular}
\end{table*}

Our calculation of the virial factor is consistent with the result presented by previous dynamical modeling studies \citep{Pancoast14b, Grier17b, Williams18, Shen24}. Based on a small sample of AGNs from the LAMP 2008 survey, \citet{Pancoast14b} reported initial measurements of the virial factor, finding  log$_{10}\,(f)_{\rm pred}=0.68\pm0.40$ using $\sigma_{\rm line, rms}$ and log$_{10}\,(f)_{\rm pred}=-0.07\pm0.40$ using FWHM$_{\rm line, mean}$. Subsequent work by \citet{Grier17b}, which incorporated five additional AGNs from the AGN10 project, derived a mean $f$ to be 0.54$\pm$0.17 with a dispersion of 0.49$\pm$0.35. Later, \citet{Williams18} updated these estimates to log$_{10}\,(f)_{\rm pred}=0.57\pm0.19$ using $\sigma_{\rm line, rms}$ and log$_{10}\,(f)_{\rm pred}=0.00\pm0.50$ using FWHM$_{\rm line, mean}$ by including seven AGNs from the LAMP 2011 campaign.  Most recently, \citet{Shen24} compiled measurements for a total of 30 AGNs, and reported a mean virial factor of 0.62$\pm$0.08 with a with a dispersion of $0.32^{+0.08}_{-0.06}$ dex. This value is consistent with our mean virial factor but the dispersion is larger. The difference can be attributed to several reasons.  First, the sample is slightly different. The inclusion of SAMP sample slightly decreases the dispersion (see Table \ref{tab:meanf_SAMP}), and \citet{Shen24} includes two additional sources modeled based on {\small{\tt BRAINS}}. Second, we utilize the full $M_{\rm BH}$ posterior samples when calculating the virial factor, whereas \citet{Shen24} assumed symmetric Gaussian distribution for $M_{\rm BH}$ based on reported median and lower and upper uncertainty. Third, the data could be slightly different. We note that Table 3 of \citet{Shen24} adopts RM virial products based on $\sigma_{\rm line,mean}$ from \citet{U22}$,$ together with other measurements derived based on $\sigma_{\rm line,rms}$. In our analysis, we use the dataset compiled by \citet{Villafana23} and Villafana et al. (2026; submitted), in which the virial products are derived consistently for each line widths and have been further updated through internal communications.

\subsection{Future prospective}

In this work, we update the estimate of virial factor and its dispersion. Equally important is understanding how the virial factor depends on various AGN properties. Based on traditional RM, it has been shown that FWHM-based virial factors exhibit significant anti-correlations with FWHM and the line shape parameter FWHM/$\sigma_{\rm line}$ \citep{Collin06, Meja-Restrepo18, Yu19, Yang24}, while virial factors based on $\sigma_{\rm line}$ display much weaker trends. These empirical relations provide empirical corrections that improve our understanding in the systematics of black hole mass estimates for large samples38. Their physical origin, however, remains not fully understood. A commonly discussed explanation invokes orientation effects: FWHM more traces a outer, disk-like component, making it more sensitive to inclination, whereas $\sigma_{\rm line}$ preferentially traces a more isotropic turbulent component and is therefore less affected by viewing angle \citep[e.g.,][]{Collin06,Shen14}. The limited knowledge on BLR geometry and kinematics from traditional RM hinder a definitive explanation of the driver of these correlations.

Based on a sample of 28 AGNs with BLR dynamical modeling, \citet{Villafana23} reported a correlation between $f$ and $M_{\rm BH}$, and confirmed the presence of correlations with the inclination $\theta_i$, the BLR opening angle $\theta_o$, and the line-shape parameter FWHM/$\sigma_{\rm line}$, although they are only marginal given the current small sample. These results support the view that the observed virial-factor trends are linked to BLR geometry and kinematics. 
As discussed in the previous section, the SAMP sample contains AGNs with relatively higher $M_{\rm BH}$ compared to earlier dynamical modeling studies. This makes it  valuable for investigating potential trends in the virial factor across a broader black hole mass range (Villafaña et al., in prep.). 

In addition to RM, near-infrared (NIR) interferometric observations based on GRAVITY, which is installed on the Very Large Telescope Interferometer (VLTI), provide an independent method for measuring black hole masses and constraining the virial factor. With its unprecedented spatial resolution, GRAVITY has successfully resolved the broad-line region (BLR) and obtained BH mass estimates for seven AGNs \citep{Gravity18, Gravity20, Gravity21, Gravity24}. The average virial factor derived from this sample is $\langle f \rangle = 3.04 \pm 0.64$, based on $\sigma_{\rm line}$ from single-epoch observations, which is consistent with our result of 3.80$\pm$1.83 ($\log_{10} f = 0.60 \pm 0.22$) derived from $\sigma_{\rm line, mean}$.

Recent improvements in sensitivity with GRAVITY+ have extended its capabilities to AGNs at cosmic noon and even beyond \citep{Abunter24, Santos25,Gravity+25}. 
Based on the single-epoch line profile of a sample of 29 AGNs at $z \sim 2$, \citet{Santos25} estimated an average $\sigma$-based H$\alpha$ virial factor of 1.44 by perform dynamical modeling. These virial factors are significantly lower than the values found for local AGNs, and they attribute this difference to the non-Gaussian shape of the emission lines that is likely resulted from outflow components \citep{Gravity+25}. 
A future direction is to perform reverberation mapping for these GRAVITY AGNs, 
which is essential for both understanding any systematic difference between the two approaches\citep[e.g.,][]{Gravity23,Gravity24b} and studying any evolution in virial factors as a function of redshifts or AGN properties.

Finally, \citet{Williams22} and \citet{Villafana24} recently introduced a second iteration of {\small\texttt{CARAMEL}}, which is referred to as {\small\texttt{CARAMEL-GAS}}. In our work, we adopt the original version of {\small\texttt{CARAMEL}}  where the BLR is modeled as a collection of point particles that instantaneously re-emit incident radiation, representing only the emissivity field. In contrast, {\small\texttt{CARAMEL-GAS}} reconstructs the gas density field of the BLR and directly links emission to the local gas density. This enables a more physically motivated framework for characterizing BLR structure and allows for the simultaneous modeling of multiple emission lines, such as H$\alpha$, H$\beta$, \ion{Mg}{2}, and \ion{C}{4}. Future studies incorporating both modeling approaches will provide deeper insights into the structure and kinematics of the BLR. In addition to these methods, several studies have explored the possibility of using single-epoch spectra \citep{Raimundo19, Raimundo20, Kuhn24}, offering a potentially very efficient approach for large AGN samples.

\section{Conclusion} \label{sec:conclusion}

We have performed dynamical modeling for eight AGNs from the SAMP to constrain their BLR geometry and kinematics and determine the BH mass. Our main results can be summarized as follows.

\begin{enumerate}
    \item Geometrically, the \hbeta-emitting broad-line region (BLR) is best described as a thick disk viewed at intermediate inclination angles. The emission predominantly originates from the far side of the BLR, with the midplane being either transparent or mildly obscured. 
    
    \item Dynamically, the BLR gas exhibits a mixture of circular, inflowing, and outflowing motions, with the inflow and outflow components contributing significantly in some cases.
    
    \item We measure the BH mass of log$_{10}(M_{\rm BH}/M_{\odot})=8.37^{+0.72}_{-0.14}$ for J0140+234,   
    $8.26^{+0.21}_{-0.16}$ for PG~0947+396, $7.72^{+0.28}_{-0.16}$ for J1026+523, $8.36^{+0.35}_{-0.25}$ for J1120+423, $8.18^{+0.29}_{-0.15}$ for PG~1121+422, 
    $7.82^{+0.24}_{-0.11}$ for J1217+333, 
    $7.46^{+0.24}_{-0.28}$ for PG~1427+480,
    $7.46^{+0.24}_{-0.16}$  for J1540+355 
    (Table \ref{tab:model_parameters_good}).

    \item By combining eight AGNs from the SAMP sample with 30 AGNs from the literature,  we determine the up-to-date mean virial factors and their uncertainties based on different types of line widths. We present the prediction of virial factors  (\(\log_{10}(f)_{\rm pred}\)) that is suitable for future RM-based BH mass calculations (Figure \ref{fig:f_distribution}). We find \(\log_{10}(f)_{\rm pred} = 0.69 \pm 0.21\) for \(\sigma_{\rm line,rms}\), \(0.58 \pm 0.25\) for \(\sigma_{\rm line,mean}\), \(0.11 \pm 0.14\) for \({\rm FWHM}_{\rm rms}\), and \(-0.08 \pm 0.23\) for \({\rm FWHM}_{\rm mean}\) (Table \ref{tab:meanf_SAMP}).

    \item The derived mean virial factors from our dynamical modeling are consistent with those obtained based on \(M_{\rm BH}\)–\(\sigma_*\) relation. This agreement supports the underlying assumption that active and inactive galaxies follow the same \(M_{\rm BH}\)–\(\sigma_*\) relation. 
    We note that the dispersion in \(\log_{10}(f)_{\rm pred}\) is only \(\sim 0.2\) dex, which is smaller than the intrinsic scatter of the AGN \(M_{\rm BH}\)–\(\sigma_*\) relation. Therefore, the updated $f$ in this work allows for a more precise of RM BH mass estimates.

    \item The SAMP sample extends the $M_{\rm BH}$ dynamic range to $M_{\rm BH}\sim 10^{8.5} M_{\odot}$.  The consistency between $f$ derived the SAMP sample and those from the combined sample suggests that similar $f$ are applicable to high-luminosity, high-mass regime, where the constraints from AGN $M_{\rm BH}$--$\sigma_*$ relation remains ambiguous. It highlights the unique advantage of dynamical modeling in constraining $M_{\rm BH}$ in these massive AGNs.

\end{enumerate}

\section{Acknowledgment}

We thank Misty Bentz and Catherine Grier for kindly sharing the {\small\texttt{CARAMEL}}  posterior samples from their previous works. We thank the anonymous referee for helpful suggestions that improved the manuscript. This work is supported by the National Research Foundation of Korea (NRF) grant funded by the Korean government (MEST) (No. 2019R1A6A1A10073437 and No. 2021R1A2C3008486).   The research at UCLA was supported by the NSF grant NSF-AST-1907208. 

\bibliography{ref}

\appendix

\section{The detailed model parameter for the eight objects included in the study}

Table \ref{tab:inputpara} lists the blue and red side wavelength range of the \hbeta\ profile as well as the intrinsic \OIII\ line width (line dispersion) used in {\small\texttt{CARAMEL}} . These information is helpful for reproducing the {\small\texttt{CARAMEL}}  results.
Table \ref{tab:model_parameters_good} lists the detailed model parameter for the eight objects. 
For completeness, in Figure 
\ref{fig:posteriorplot} we also present the 2-dimensional (2-D) posterior distribution for the most critical parameters, i.e, black hole mass (log$_{10}M_{\rm BH}$), median  \hbeta\ BLR radius ($r_{\rm median}$),  inclination angle ($\theta_i$), and opening angle ($\theta_o$). As can be seen, there is an anti-correlation between the black hole mass and inclination angle, which is expected given their dependence on the observed line width.

\renewcommand{\thetable}{A\arabic{table}}
\setcounter{table}{0} 
\begin{table}[htbp]
\centering
\caption{Input H$\beta$ Profile Range and [O\,\textsc{iii}] Line Width Used in the CARAMEL Modeling} \label{tab:inputpara}
\begin{tabular}{lcc}
\hline
Object & H$\beta$ Profile Range (\AA) & [O\,\textsc{iii}] $\sigma_{\rm line}$ (km\,s$^{-1}$) \\
\hline
J0140+234 & 4750--4975 & 234 \\
PG0947+396 & 4760--4960 & 238 \\
J1026+523 & 4760--4960 & 205 \\
J1120+423 & 4760--4960 & 296 \\
PG1121+422 & 4760--4960 & 224 \\
J1217+333 & 4760--4960 & 214 \\
PG1427+480 & 4760--4960 & 282 \\
J1540+355 & 4760--4960 & 247 \\
\hline
\end{tabular}
\end{table}

\begin{sidewaystable}
\centering
\caption{BLR Model Parameters}
\begin{tabular}{lcccccccc}
\hline
Quantity & J0140+234 & PG0947+396 & J1026+523 & J1120+423 & PG1121+422 & J1217+333 & PG1427+480 & J1540+355 \\
\hline
$R_{\rm mean}$ & $135_{-12}^{+18}$ & $55.7_{-7.0}^{+7.3}$ & $40.7_{-6.1}^{+9.1}$ & $109_{-63}^{+35}$ & $111.4_{-6.6}^{+13}$ & $55.9_{-6.9}^{+4.6}$ & $46_{-21}^{+25}$ & $80_{-22}^{+29}$ \\
$R_{\rm median}$ & $91_{-23}^{+24}$ & $47.5_{-6.2}^{+6.0}$ & $27.6_{-5.0}^{+5.9}$ & $81_{-43}^{+36}$ & $83.5_{-5.8}^{+8.0}$ & $38.5_{-6.6}^{+4.9}$ & $25_{-13}^{+11}$ & $39_{-13}^{+12}$ \\
$R_{\rm min}$ & $28.9_{-7.9}^{+6.9}$ & $11.9_{-8.5}^{+11}$ & $6.4_{-4.2}^{+5.0}$ & $18_{-17}^{+21}$ & $23.1_{-9.1}^{+13}$ & $13.3_{-5.6}^{+6.8}$ & $5.8_{-3.7}^{+3.3}$ & $7.0_{-2.3}^{+6.3}$ \\
$\sigma_r$ & $197_{-69}^{+320}$ & $32.9_{-6.5}^{+7.4}$ & $39.4_{-7.4}^{+12}$ & $88_{-58}^{+30}$ & $99_{-16}^{+21}$ & $47.9_{-6.3}^{+10}$ & $63_{-32}^{+37}$ & $132_{-49}^{+260}$ \\
$\tau_{\rm mean}$ & $133_{-18}^{+21}$ & $53.7_{-6.5}^{+6.4}$ & $44.6_{-6.1}^{+9.2}$ & $73.7_{-37}^{+18}$ & $112.0_{-8.2}^{+8.0}$ & $61.6_{-4.6}^{+5.2}$ & $41_{-17}^{+20}$ & $82_{-23}^{+28}$ \\
$\tau_{\rm median}$ & $86_{-25}^{+22}$ & $42.7_{-5.9}^{+6.0}$ & $27.5_{-5.0}^{+5.7}$ & $44.3_{-21}^{+8.3}$ & $77.3_{-4.7}^{+6.7}$ & $39.0_{-4.6}^{+5.2}$ & $19.3_{-9.5}^{+8.7}$ & $34_{-11}^{+12}$ \\
$\beta$ & $1.31_{-0.29}^{+0.45}$ & $0.76_{-0.17}^{+0.24}$ & $1.16_{-0.20}^{+0.23}$ & $0.93_{-0.20}^{+0.23}$ & $1.05_{-0.17}^{+0.20}$ & $1.19_{-0.19}^{+0.25}$ & $1.44_{-0.16}^{+0.14}$ & $1.58_{-0.14}^{+0.16}$ \\
$\theta_o$ & $27.2_{-7.8}^{+11}$ & $29.5_{-7.5}^{+9.1}$ & $44_{-14}^{+12}$ & $67.8_{-6.5}^{+13}$ & $29.4_{-8.0}^{+7.5}$ & $63_{-17}^{+22}$ & $57_{-17}^{+15}$ & $47.2_{-8.6}^{+14}$ \\
$\theta_i$ & $21.7_{-8.8}^{+9.8}$ & $28.1_{-6.5}^{+8.5}$ & $29.7_{-12}^{+9.1}$ & $40.8_{-8.4}^{+14}$ & $24.7_{-8.0}^{+8.3}$ & $31.6_{-10}^{+6.1}$ & $25.3_{-9.4}^{+19}$ & $45.1_{-18}^{+7.5}$ \\
$\kappa$ & $-0.43_{-0.06}^{+0.70}$ & $-0.26_{-0.15}^{+0.13}$ & $-0.32_{-0.12}^{+0.08}$ & $0.16_{-0.28}^{+0.22}$ & $-0.43_{-0.06}^{+0.20}$ & $-0.44_{-0.03}^{+0.14}$ & $-0.49_{-0.01}^{+0.05}$ & $-0.45_{-0.04}^{+0.09}$ \\
$\gamma$ & $1.76_{-0.25}^{+0.17}$ & $1.59_{-0.32}^{+0.29}$ & $1.49_{-0.33}^{+0.34}$ & $1.66_{-0.38}^{+0.28}$ & $1.77_{-0.34}^{+0.17}$ & $1.36_{-0.24}^{+0.39}$ & $1.56_{-0.45}^{+0.35}$ & $1.22_{-0.16}^{+0.22}$ \\
$\xi$ & $0.49_{-0.22}^{+0.27}$ & $0.50_{-0.12}^{+0.15}$ & $0.83_{-0.20}^{+0.12}$ & $0.59_{-0.32}^{+0.17}$ & $0.51_{-0.19}^{+0.28}$ & $0.66_{-0.16}^{+0.26}$ & $0.17_{-0.06}^{+0.08}$ & $0.17_{-0.13}^{+0.22}$ \\
$M_{\rm BH}$ & $8.37_{-0.14}^{+0.72}$ & $8.26_{-0.16}^{+0.21}$ & $7.72_{-0.16}^{+0.28}$ & $8.36_{-0.25}^{+0.35}$ & $8.18_{-0.15}^{+0.29}$ & $7.82_{-0.11}^{+0.24}$ & $7.46_{-0.28}^{+0.24}$ & $7.46_{-0.16}^{+0.24}$ \\
$f_{\rm ellip}$ & $0.39_{-0.18}^{+0.37}$ & $0.28_{-0.13}^{+0.10}$ & $0.29_{-0.16}^{+0.16}$ & $0.51_{-0.37}^{+0.17}$ & $0.36_{-0.10}^{+0.09}$ & $0.40_{-0.29}^{+0.25}$ & $0.05_{-0.04}^{+0.06}$ & $0.24_{-0.14}^{+0.23}$ \\
$f_{\rm flow}$ & $0.34_{-0.23}^{+0.40}$ & $0.26_{-0.17}^{+0.17}$ & $0.74_{-0.18}^{+0.18}$ & $0.67_{-0.17}^{+0.22}$ & $0.26_{-0.18}^{+0.15}$ & $0.18_{-0.12}^{+0.23}$ & $0.31_{-0.19}^{+0.30}$ & $0.66_{-0.28}^{+0.22}$ \\
$\theta_e$ & $9.7_{-6.6}^{+10}$ & $24_{-12}^{+14}$ & $24_{-15}^{+19}$ & $48_{-30}^{+23}$ & $10.6_{-7.5}^{+7.8}$ & $14.0_{-9.0}^{+10}$ & $7.9_{-5.2}^{+4.8}$ & $14.0_{-8.9}^{+11}$ \\
In.$-$Out. & $-0.60$ & $-0.66$ & $0.65$ & $0.33$ & $-0.61$ & $-0.58$ & $-0.94$ & $0.74$ \\
$\sigma_{\rm turb}$ & $0.02_{-0.02}^{+0.05}$ & $0.01_{-0.01}^{+0.03}$ & $0.01_{-0.01}^{+0.04}$ & $0.01_{-0.01}^{+0.03}$ & $0.06_{-0.05}^{+0.03}$ & $0.02_{-0.02}^{+0.05}$ & $0.01_{-0.01}^{+0.03}$ & $0.02_{-0.01}^{+0.05}$ \\
\hline
\end{tabular}
\end{sidewaystable}
 \label{tab:model_parameters_good}

\section{Dynamical modeling results for excluded objects} \label{sec:AppendixA}

Based on our inspection on the fitting to the light curve as well as the posterior distribution, we selected eight objects that have reasonably good fitting quality. This section presents the fitting results for the seven excluded objects as shown in  Figure \ref{fig:Example_Fitting_B}. We provide brief description for these objects in the following.

For J0101+422, the model fit shows noticeable deviations in the second-year data, yielding a relatively large reduced $\chi^2$ of 3.45. The primary reason for excluding this object is that its posterior samples did not converge, indicating that 
{\small\texttt{CARAMEL}} struggled to find a stable solution.
For PG~1202+281, the model fits display significant deviations during the first, second, and fourth years, yielding a relatively large reduced $\chi^2$ of 3.60. Its $M_{\rm BH}$ posterior distribution also exhibits a double-peaked profile. For these reasons, we exclude this object from our analysis.
For VIII~Zw~218, the model fit shows a very significant deviation in the third-year data, although the reduced $\chi^2$ is only 1.49. This is because the statistical temperature $T$ is usually high as reflected in the error bars of the residual plots. For this object, the $T$ suggests that the model is unable to reproduce the data given reasonable uncertainties. 
For PG~1322+659, the variability is not strong and the model tends to be flat. In addition,  the posterior distribution does not appear to be well converged so we decide to exclude this object.
For PG~1440+356, the fits requires a a large $T$ value and obtains a very small reduced $\chi^2$ of 0.07, indicating the model overfits the data. In addition,  we identified inconsistencies in the model’s error bars across different velocity bins. Hence, we exclude this object.
For J1456+380 and J1619+501, the main reason for exclusion is that {\small\texttt{CARAMEL}}  did not achieve satisfactory convergence.

\renewcommand{\thefigure}{A\arabic{figure}}
\setcounter{figure}{0} 
\begin{figure*}[htbp]
    \includegraphics[width=0.5\textwidth]{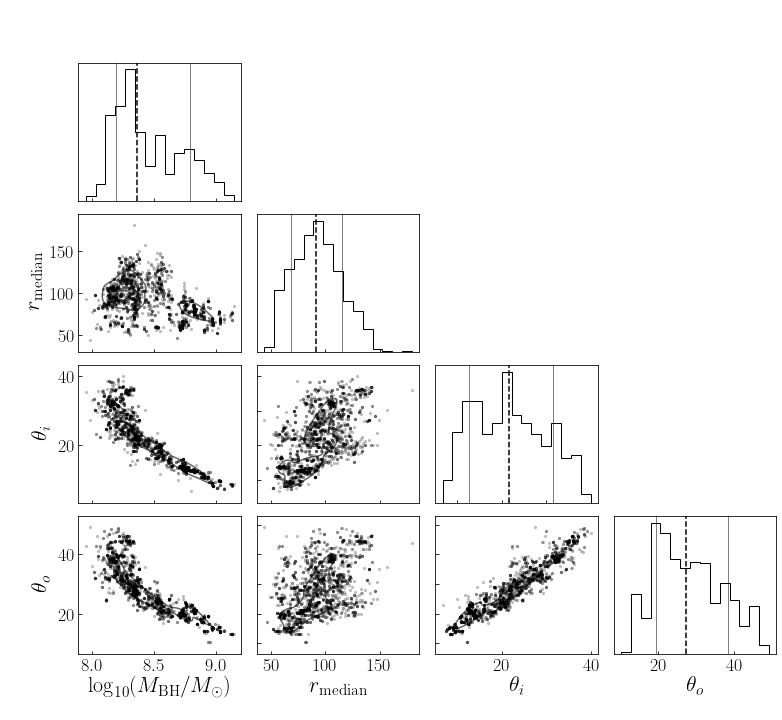}
    \includegraphics[width=0.5\textwidth]{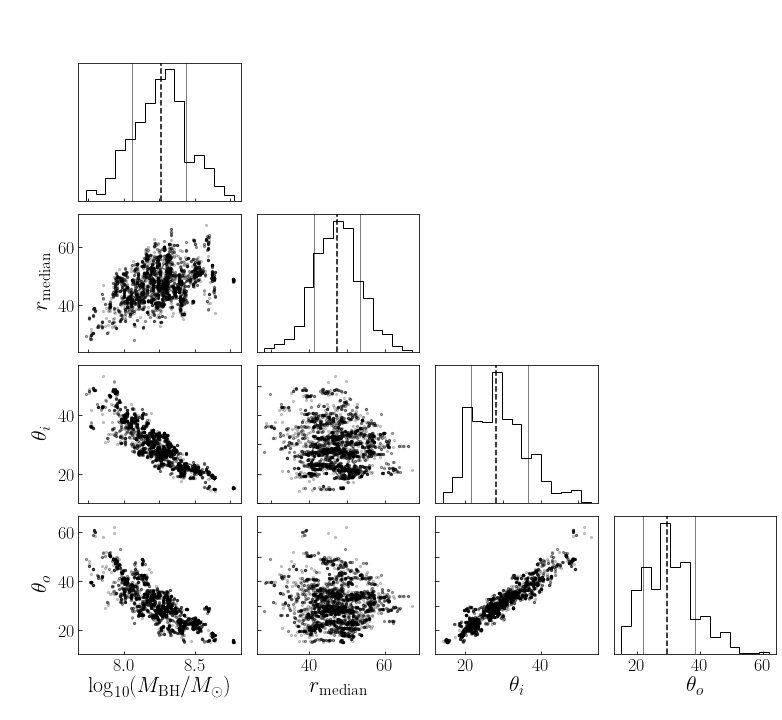}

    \includegraphics[width=0.5\textwidth]{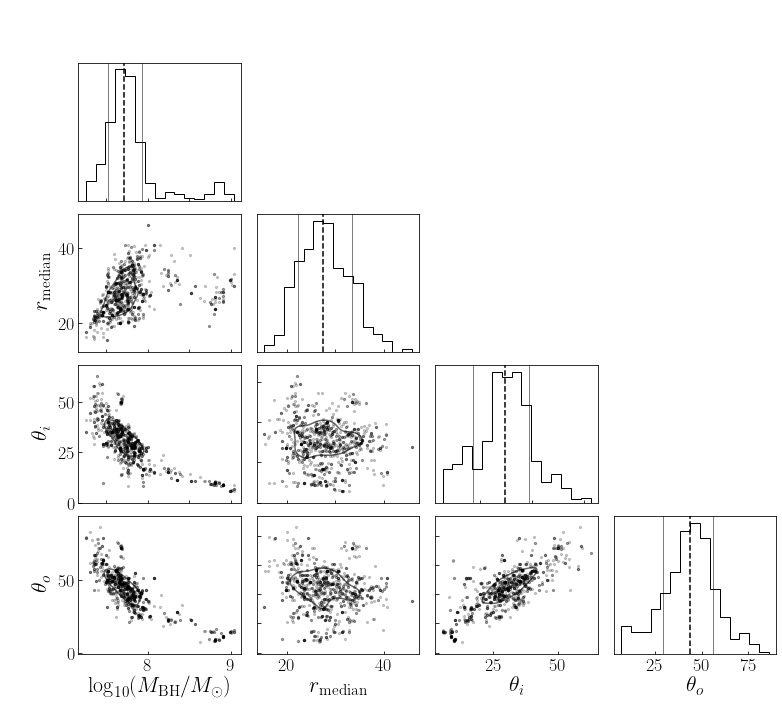}
    \includegraphics[width=0.5\textwidth]{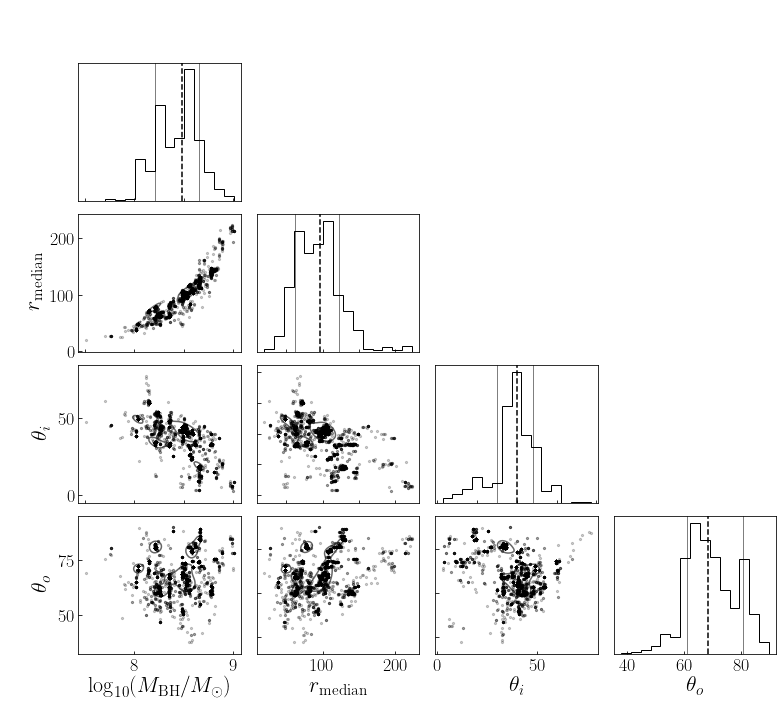}

    \caption{2D posterior distributions for the black hole mass (log$_{10}M_{\rm BH}$), median 
    \hbeta\ BLR radius ($r_{\rm median}$),  inclination angle ($
    \theta_i$), and opening angle ($\theta_o$). The median and 68\% confidence intervals are indicated by the vertical dashed and dotted lines, respectively. From top left to bottom right, the panels correspond to J0140+234, PG~0947+356, J1026+523, and J1120+423.} \label{fig:posteriorplot}
\end{figure*}

\setcounter{figure}{0} 
\begin{figure*}[htbp]

    \includegraphics[width=0.5\textwidth]{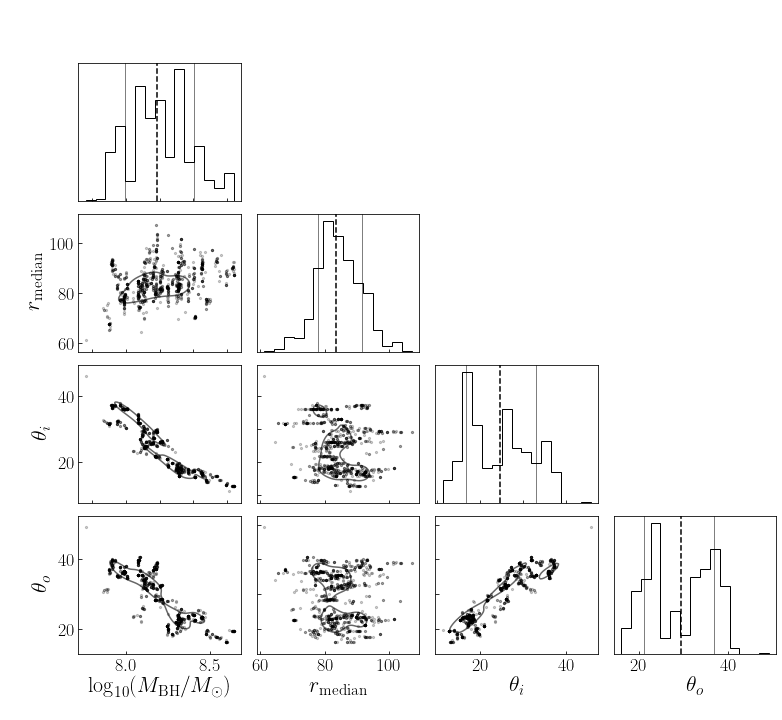}
    \includegraphics[width=0.5\textwidth]{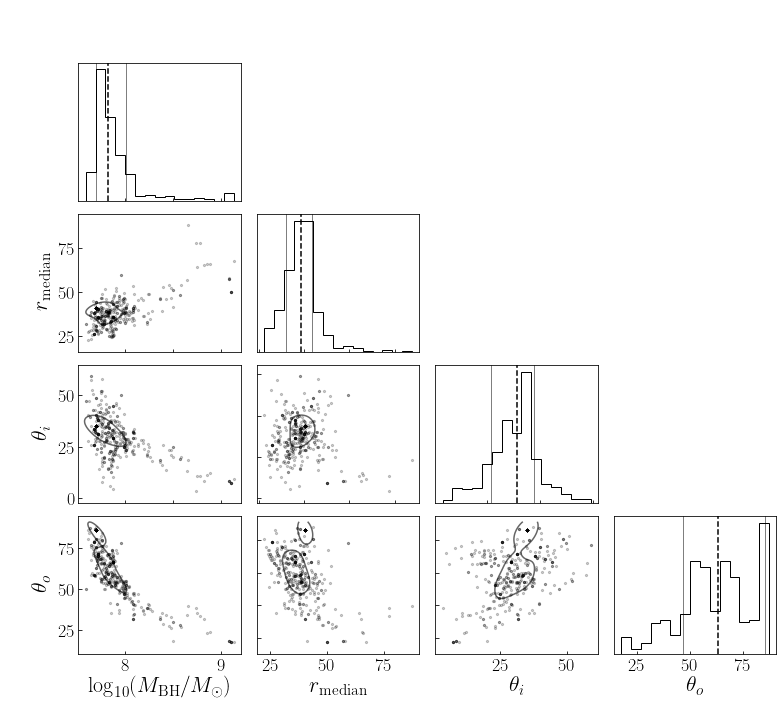}

    \includegraphics[width=0.5\textwidth]{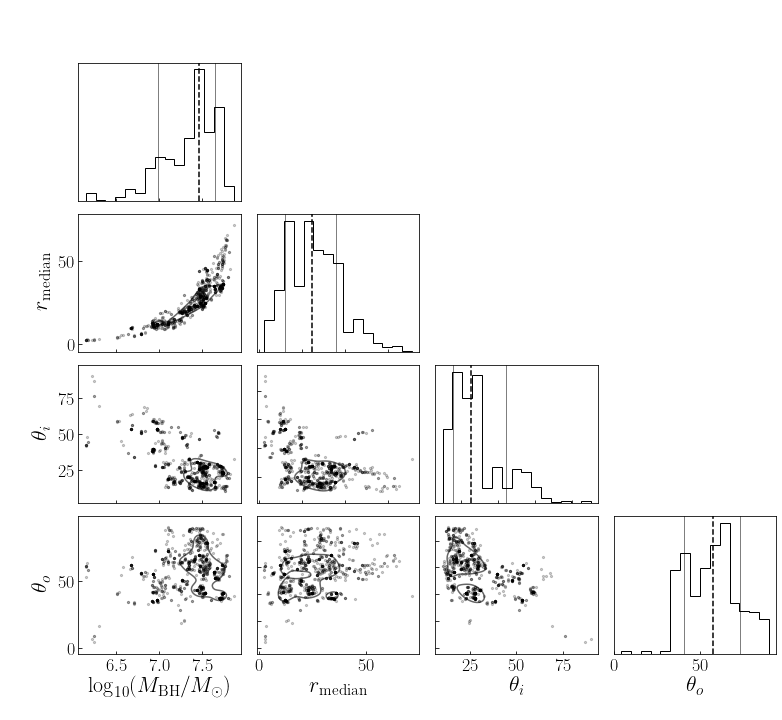}
    \includegraphics[width=0.5\textwidth]{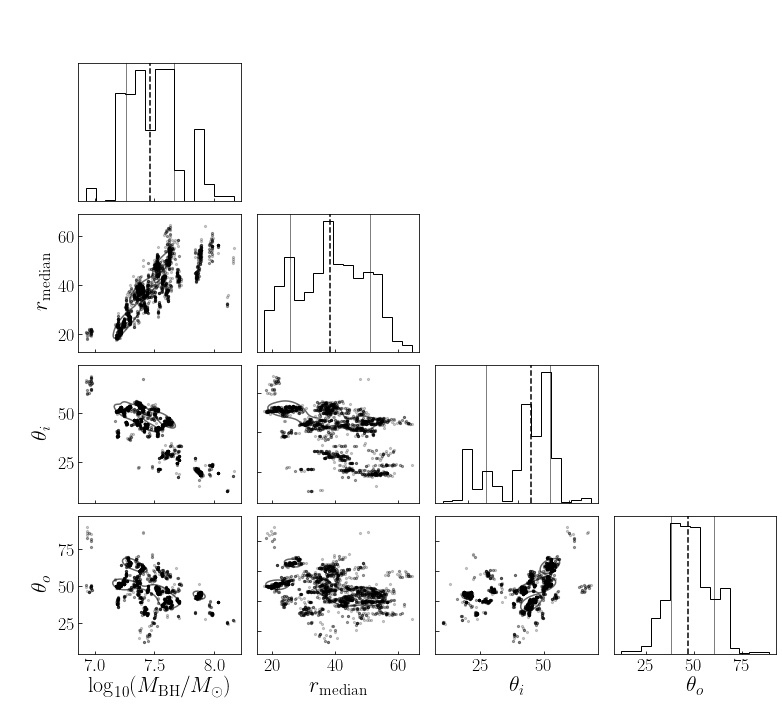}
    \caption{Continued. From top left to bottom right, the panels correspond to PG~1121+422, J1217+333, PG~1427+480, and J1540+355.} 
\end{figure*}

\renewcommand{\thefigure}{B\arabic{figure}}
\setcounter{figure}{0} 
\begin{figure*}
    \centering

\includegraphics[width=0.49\textwidth]{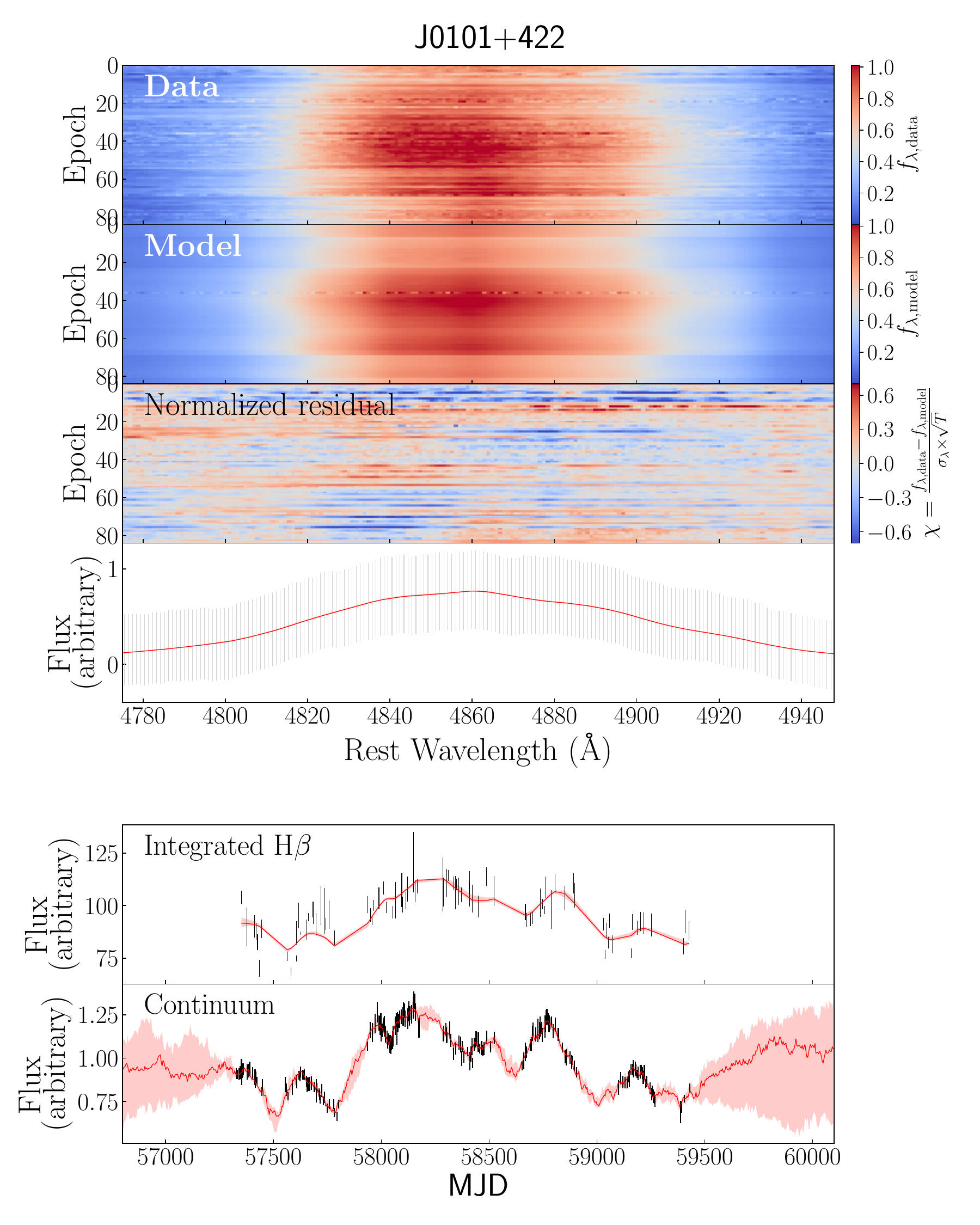}
\includegraphics[width=0.49\textwidth]{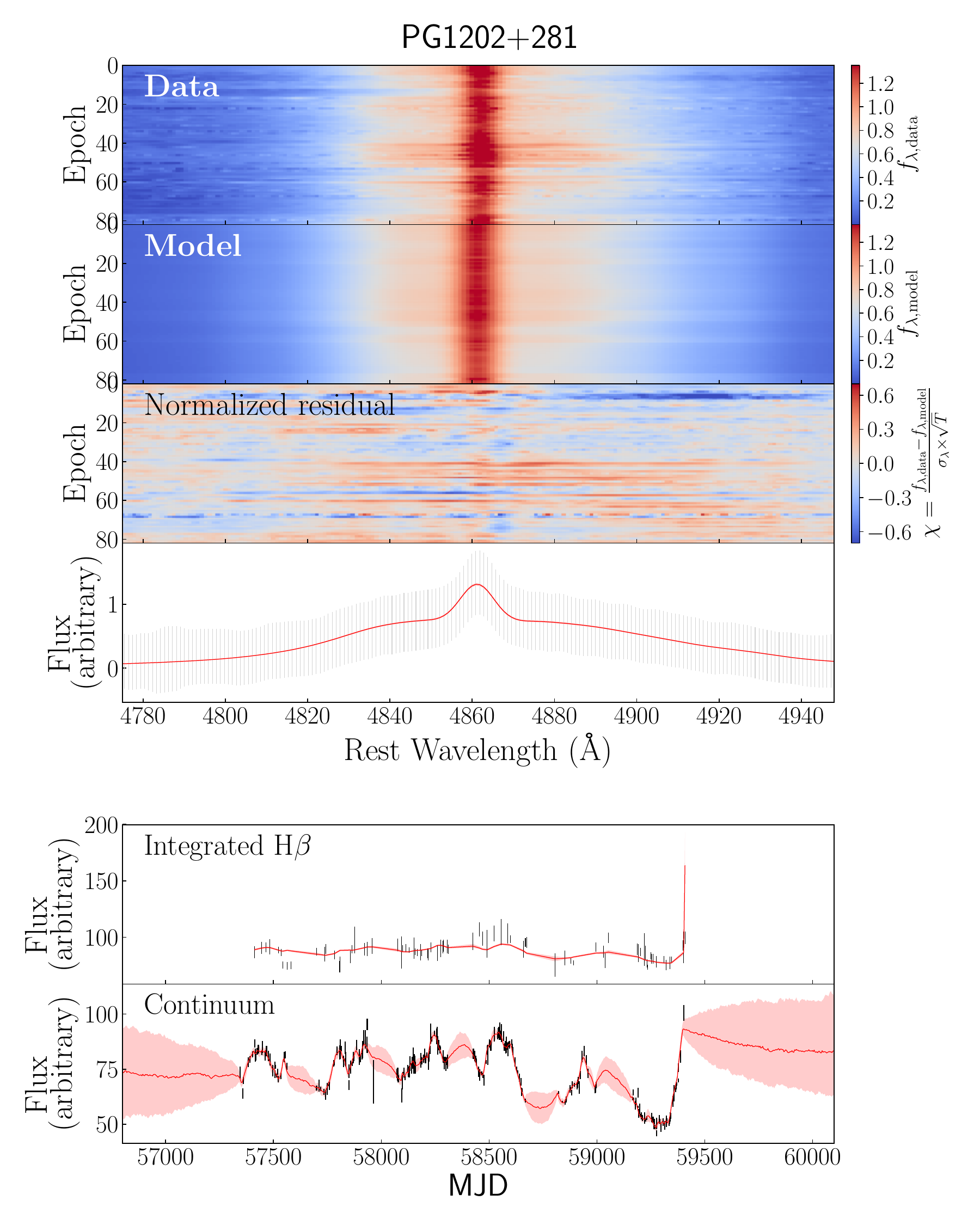}
\includegraphics[width=0.49\textwidth]{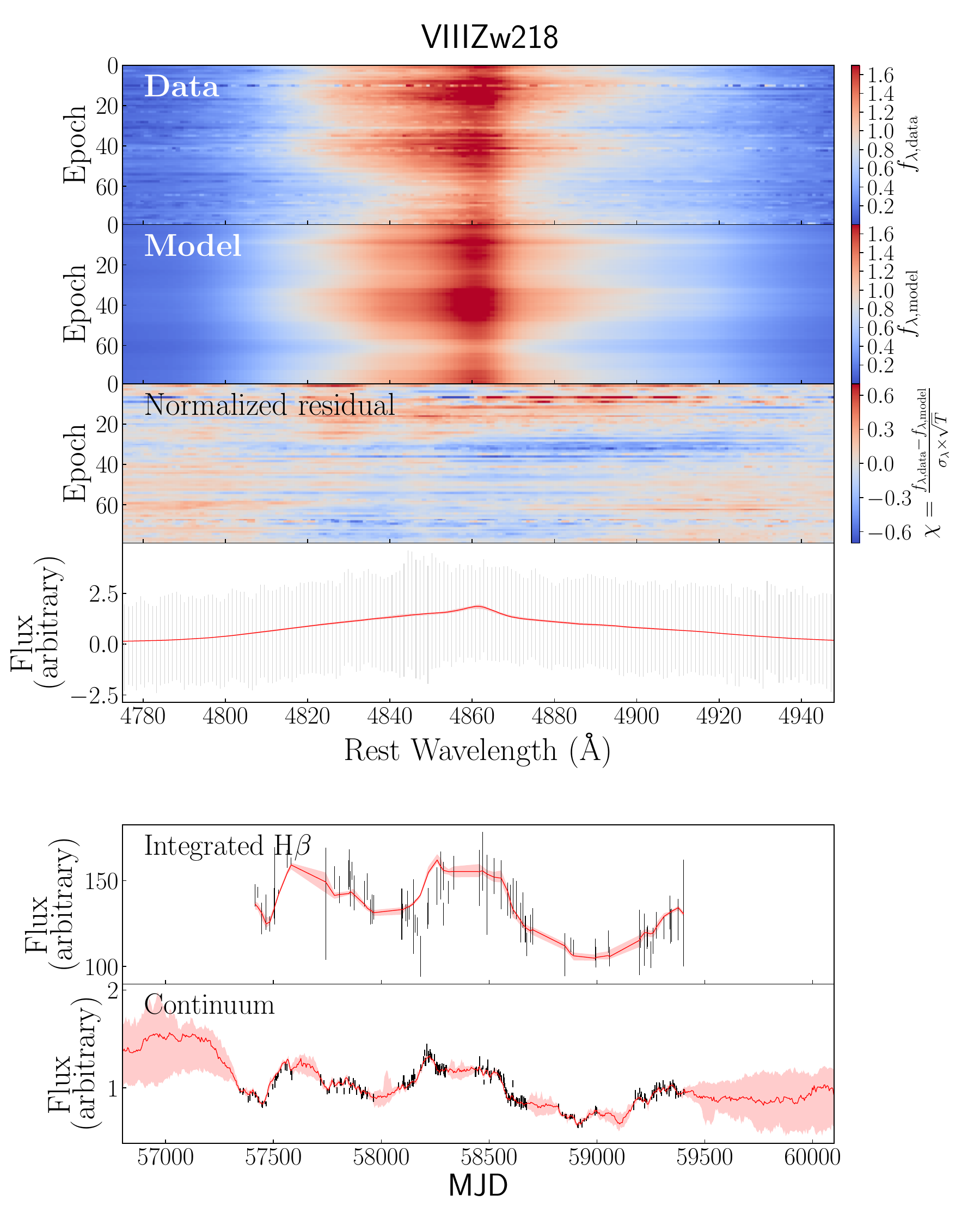}
\includegraphics[width=0.49\textwidth]{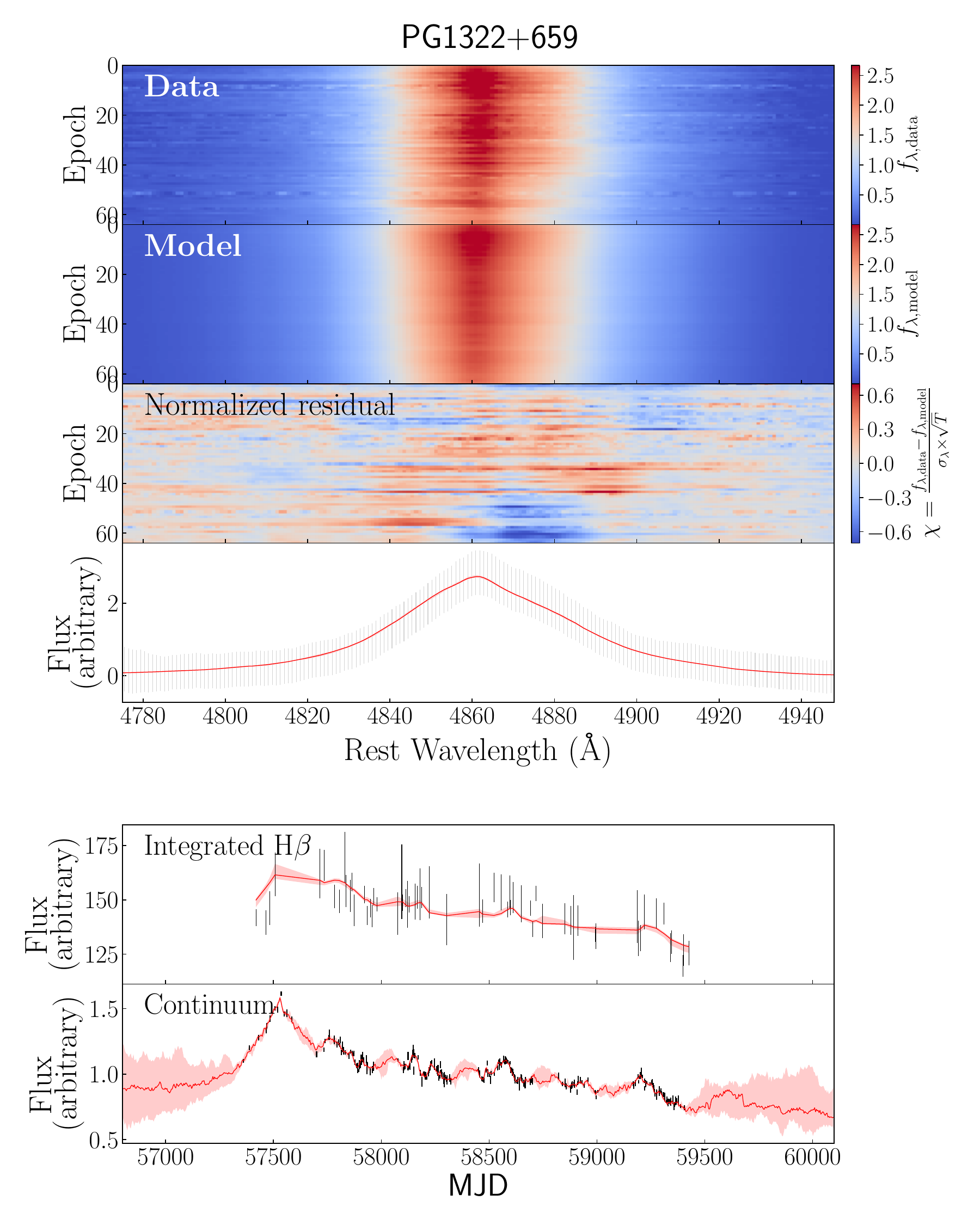}

    \caption{Same as Figure \ref{fig:Example_Fitting} but for excluded objects, including J0101+422, PG1202+281, VIII~Zw~218, PG1322+659, PG~1440+356, J1456+380, and J1619+501.} \label{fig:Example_Fitting_B}
\end{figure*}

\addtocounter{figure}{-1}
\begin{figure*}
    \centering

\includegraphics[width=0.49\textwidth]{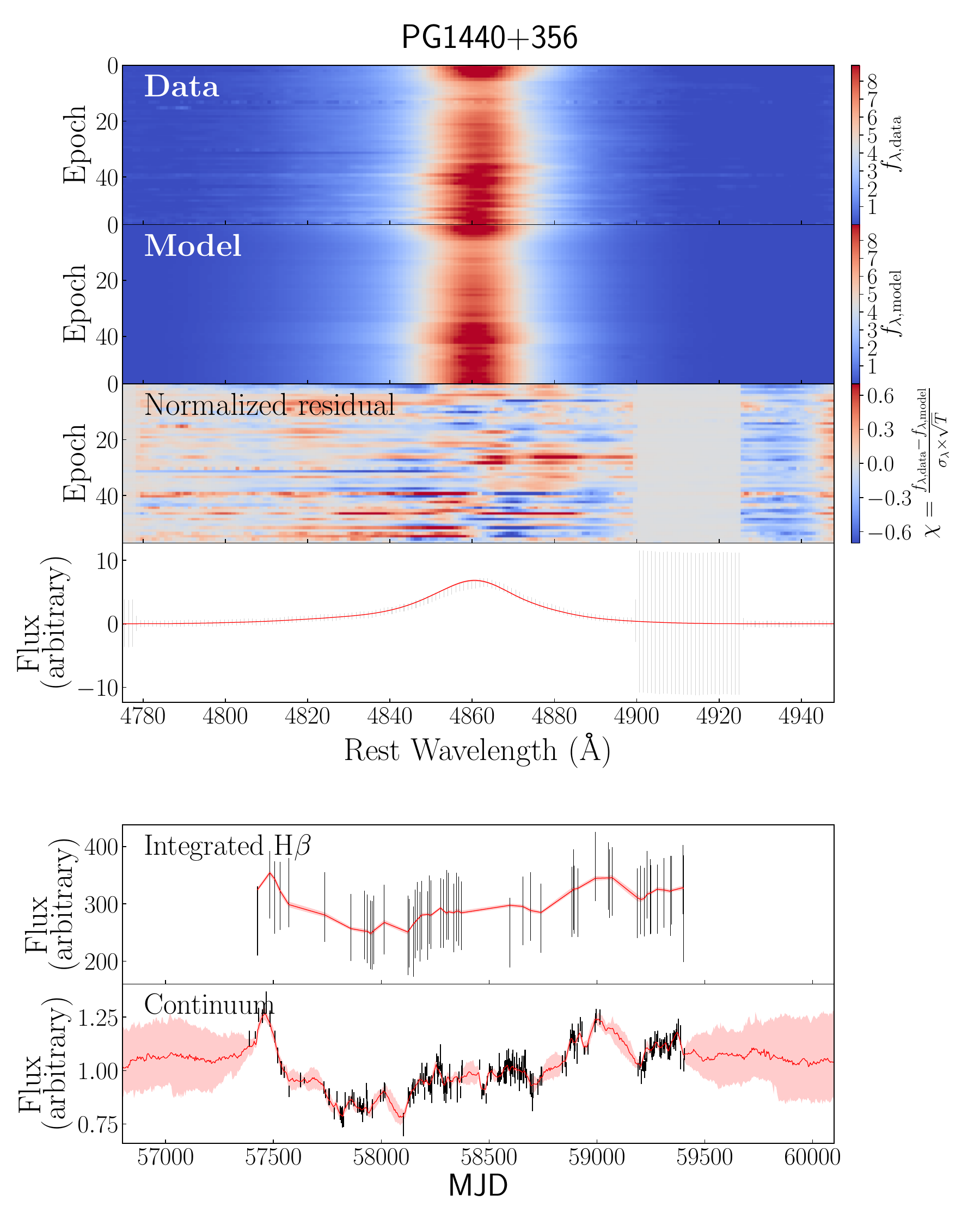}
\includegraphics[width=0.49\textwidth]{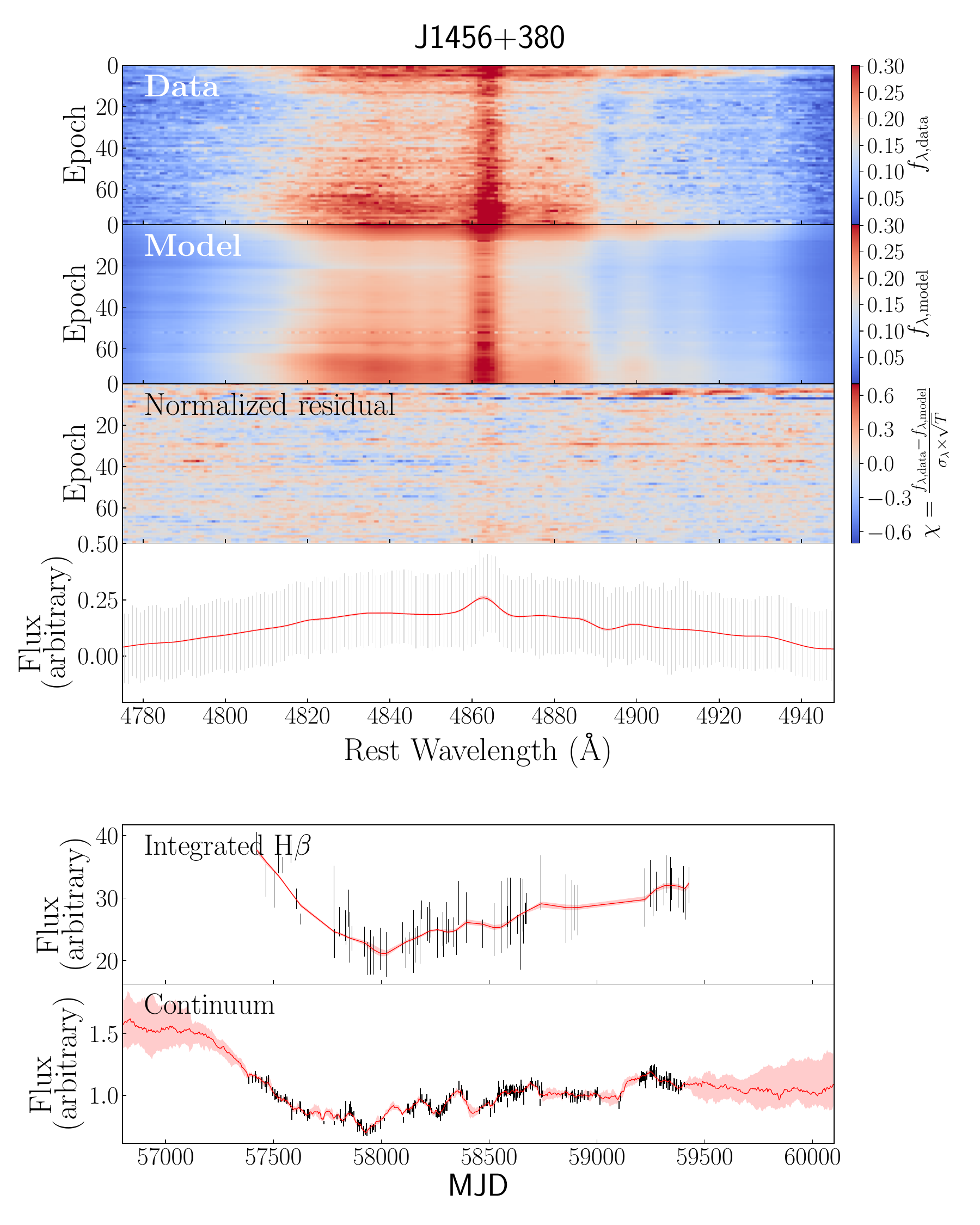}
\includegraphics[width=0.49\textwidth]{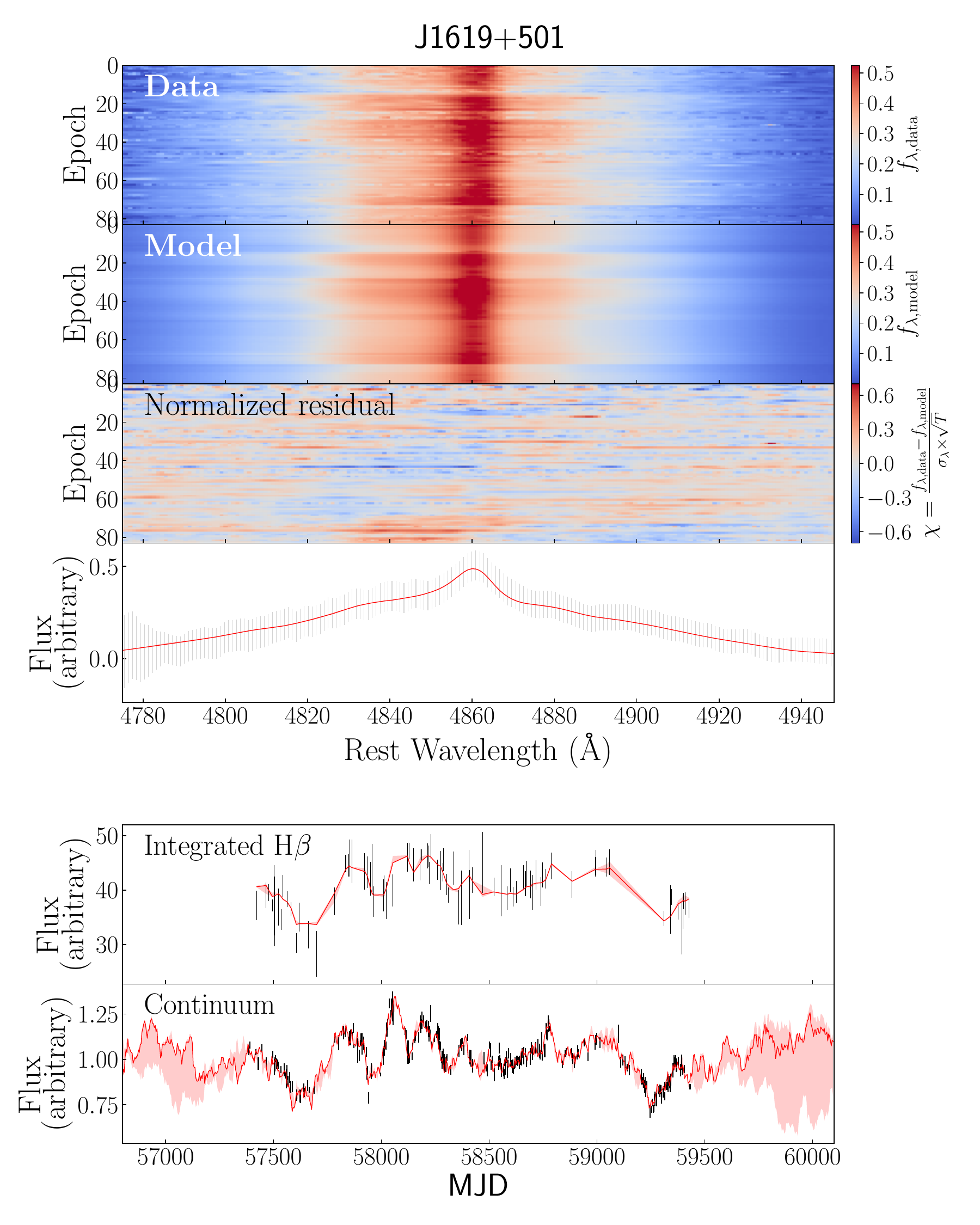}
    \caption{Continued. }
\end{figure*}

\section{Virial Factors for AGNs in the Literature Sample} \label{sec:AppendixB}

This sections lists the virial factors for each individual AGN in the literature sample (Table \ref{tab:individual_f_lit}). 

\renewcommand{\thetable}{C\arabic{table}}
\setcounter{table}{0}
\begin{table*}
\centering
\caption{Virial Factors for Each Individual AGN in the Literature Sample.}
\label{tab:individual_f_lit}
 \begin{tabular}{lcccc c}
\hline
Object & logf$_{\rm \sigma,rms}$ & logf$_{\rm FWHM,rms}$ & logf$_{\rm \sigma,mean}$ & logf$_{\rm FWHM,mean}$ & References \\ \hline 

Arp~151 & $0.51_{-0.14}^{+0.13}$ & $-0.04_{-0.14}^{+0.12}$ & $0.27_{-0.14}^{+0.12}$ & $-0.24_{-0.14}^{+0.12}$ & (1), (2), (3) \\
Mrk~1310 & $1.67_{-0.32}^{+0.27}$ & $1.06_{-0.28}^{+0.27}$ & $1.41_{-0.28}^{+0.26}$ & $0.82_{-0.28}^{+0.26}$ & (1), (2), (3) \\
NGC~5548$^a$ & $0.46_{-0.20}^{+0.29}$ & $-0.52_{-0.26}^{+0.30}$ & $0.37_{-0.21}^{+0.28}$ & $-0.54_{-0.22}^{+0.27}$  & (1), (2), (3)\\
NGC~6814 & $-0.13_{-0.23}^{+0.30}$ & $-0.63_{-0.19}^{+0.28}$ & $-0.17_{-0.18}^{+0.25}$ & $-0.68_{-0.19}^{+0.25}$  & (1), (2), (3)\\
SBS~1116+583A & $1.01_{-0.29}^{+0.33}$ & $0.30_{-0.40}^{+0.45}$ & $1.04_{-0.29}^{+0.32}$ & $0.37_{-0.28}^{+0.32}$  & (1), (2), (3) \\
Mrk~335 & $0.63_{-0.12}^{+0.12}$ & $0.28_{-0.11}^{+0.11}$ & $0.55_{-0.10}^{+0.11}$ & $0.20_{-0.10}^{+0.11}$ & (1), (4), (5)\\
Mrk~1501 & $1.22_{-0.21}^{+0.27}$ & $0.44_{-0.21}^{+0.27}$ & $1.17_{-0.21}^{+0.26}$ & $0.36_{-0.21}^{+0.27}$ & (1), (4), (5) \\
3C~120 & $0.97_{-0.20}^{+0.16}$ & $0.53_{-0.20}^{+0.15}$ & $1.01_{-0.19}^{+0.15}$ & $0.22_{-0.20}^{+0.15}$ & (1), (4), (5) \\
PG~2130+099 & $0.34_{-0.21}^{+0.21}$ & $0.37_{-0.21}^{+0.22}$ & $0.42_{-0.20}^{+0.20}$ & $0.02_{-0.19}^{+0.20}$ & (1), (4), (5) \\
Mrk~50 & $0.68_{-0.21}^{+0.25}$ & $0.24_{-0.20}^{+0.26}$ & $0.68_{-0.20}^{+0.25}$ & $0.06_{-0.20}^{+0.26}$ & (1), (6), (7)\\
Mrk~279 & $0.66_{-0.11}^{+0.12}$ & $0.13_{-0.14}^{+0.15}$ & $0.64_{-0.11}^{+0.11}$ & $-0.07_{-0.10}^{+0.11}$ & (1), (6), (7)\\
Mrk~1511 & $0.76_{-0.18}^{+0.21}$ & $0.09_{-0.18}^{+0.21}$ & $0.59_{-0.18}^{+0.21}$ & $-0.12_{-0.18}^{+0.21}$ & (1), (6), (7)\\
NGC~4593 & $0.43_{-0.19}^{+0.28}$ & $-0.27_{-0.19}^{+0.28}$ & $0.27_{-0.19}^{+0.28}$ & $-0.42_{-0.19}^{+0.28}$ & (1), (6), (7)\\
Zw~229-015 & $0.45_{-0.28}^{+0.41}$ & $0.36_{-0.27}^{+0.40}$ & $0.38_{-0.27}^{+0.41}$ & $-0.28_{-0.28}^{+0.41}$ & (1), (6), (7)\\
NGC~5548$^b$ & $0.50_{-0.28}^{+0.32}$ & $-0.50_{-0.32}^{+0.35}$ & $0.40_{-0.28}^{+0.31}$ & $-0.51_{-0.28}^{+0.31}$ & (1), (8)\\
PG~2209+184 & $0.87_{-0.20}^{+0.21}$ & $0.11_{-0.20}^{+0.21}$ & $0.74_{-0.20}^{+0.21}$ & $-0.08_{-0.20}^{+0.20}$ & (1), (9), (10)\\
RBS~1917 & $0.87_{-0.35}^{+0.31}$ & $0.29_{-0.34}^{+0.33}$ & $0.57_{-0.32}^{+0.29}$ & $-0.05_{-0.31}^{+0.28}$  & (1), (9), (10)\\
MCG+04-22-042 & $1.23_{-0.31}^{+0.40}$ & $0.55_{-0.31}^{+0.39}$ & $1.09_{-0.31}^{+0.40}$ & $0.36_{-0.31}^{+0.39}$  & (1), (9), (10)\\
NPM1G~+27.0587 & $1.04_{-0.49}^{+0.50}$ & $0.58_{-0.44}^{+0.52}$ & $1.05_{-0.45}^{+0.51}$ & $0.42_{-0.46}^{+0.52}$  & (1), (9), (10)\\
Mrk~1392 & $1.11_{-0.12}^{+0.13}$ & $0.33_{-0.13}^{+0.13}$ & $1.04_{-0.13}^{+0.13}$ & $0.21_{-0.13}^{+0.13}$   & (1), (9), (10)\\
RBS~1303 & $0.05_{-0.19}^{+0.27}$ & $-0.21_{-0.16}^{+0.24}$ & $0.08_{-0.15}^{+0.24}$ & $-0.46_{-0.15}^{+0.24}$  & (1), (9), (10)\\
Mrk~1048 & $1.08_{-0.59}^{+0.62}$ & $0.33_{-0.57}^{+0.64}$ & $1.02_{-0.58}^{+0.62}$ & $0.18_{-0.58}^{+0.63}$  & (1), (9), (10)\\
RXJ~2044.0+2833 & $0.78_{-0.21}^{+0.18}$ & $0.04_{-0.19}^{+0.18}$ & $0.68_{-0.21}^{+0.18}$ & $-0.02_{-0.19}^{+0.17}$  & (1), (9), (10)\\
Mrk~841 & $0.65_{-0.36}^{+0.49}$ & $-0.38_{-0.36}^{+0.49}$ & $0.70_{-0.36}^{+0.50}$ & $-0.34_{-0.36}^{+0.51}$  & (1), (9), (10) \\
NGC~3783 & $0.52_{-0.13}^{+0.14}$ & $0.03_{-0.19}^{+0.20}$ & $0.50_{-0.13}^{+0.14}$ & $-0.37_{-0.13}^{+0.15}$ & (1), (11)\\
NGC~4151 & $0.52_{-0.13}^{+0.14}$ & $0.03_{-0.19}^{+0.20}$ & $0.50_{-0.13}^{+0.14}$ & $-0.37_{-0.13}^{+0.15}$ & (1), (12)\\
NGC~3227 & $0.75_{-0.36}^{+0.35}$ & $0.06_{-0.36}^{+0.35}$ & $0.62_{-0.35}^{+0.36}$ & $-0.21_{-0.35}^{+0.35}$ & (13)\\
IC~4329A & $0.49_{-0.27}^{+0.55}$ & $-0.18_{-0.31}^{+0.52}$ & $0.44_{-0.26}^{+0.55}$ & $-0.54_{-0.26}^{+0.55}$ &  (14)\\
\hline
\multicolumn{6}{l}{\parbox{13cm}{ Notes. Virial factors of individual objects for the sample of 30 AGNs from literature sample, including 28 AGNs compiled by \citet{Villafana23} and 2 additional AGNs from \citet{Bentz23a} and \citet{Bentz23b}.
Notes (a) and (b): NGC 5548 has two independent dynamical modeling measurements; the first row corresponds to the LAMP 2008 campaign \citep{Bentz09b, Pancoast14b}, while the second is based on the STORM campaign \citep{Williams20}.

References:
(1) \citet{Villafana23}; (2) \citet{Bentz09b}; (3) \citet{Pancoast14b}; (4) \citet{Grier13a}; (5) \citet{Grier17b}; (6) \citet{Barth15}; (7) \citet{Williams18}; (8) \citet{Williams20}; (9) \citet{U22}; (10) \citet{Villafana22}; (11) \citet{Bentz21}; (12) \citet{Bentz22}; (13) \citet{Bentz23a}; (14) \citet{Bentz23b}.}}
\end{tabular}
\end{table*}

\end{document}